# Learning Stationary Correlated Equilibria in Constrained General-Sum Stochastic Games

Vesal Hakami and Mehdi Dehghan

*Abstract*—We study constrained general-sum stochastic games with unknown Markovian dynamics. A distributed constrained no-regret $Q$-learning scheme (CNR$Q$) is presented to guarantee convergence to the set of stationary correlated equilibria of the game. Prior art addresses the unconstrained case only, is structured with nested control loops, and has no convergence result. CNR$Q$ is cast as a single-loop three-timescale asynchronous stochastic approximation algorithm with set-valued update increments. A rigorous convergence analysis with differential inclusion arguments is given which draws on recent extensions of the theory of stochastic approximation to the case of asynchronous recursive inclusions with set-valued mean fields. Numerical results are given for the exemplary application of CNR$Q$ to decentralized resource control in heterogeneous wireless networks (HetNets).

## I. INTRODUCTION

### A. Research Background

Stochastic games [1] are very broad framework, generalizing both Markov decision processes (MDPs) and repeated games. In particular, stochastic games are extensions of MDPs to the multi-agent case, and of repeated games to the multi-state case. A stochastic game is played in a sequence of stages. At the beginning of each stage, the game is in a certain state. The agents select their actions, and each agent receives a reward that depends on both the current state and the action profile of all the agents. The game then transitions to a new state with a certain probability which, by Markov property, depends only on the previous state and the actions chosen by all agents. This process recurs at the new state, and the interaction goes on for a finite or infinite number of stages. Similarly to the case with MDPs, each agent participating in a stochastic game aims to maximize an expected cumulative reward measure often calculated as either average reward per stage or total discounted reward. However, the solution concept differs from the case of MDPs in that the agents should settle instead for competitive optimality which corresponds to some notion of strategic equilibrium. The most common notions of equilibria are Nash [2] and correlated equilibria [3]. A Nash equilibrium (NE) is a vector of independent strategies, each of which is a probability distribution over actions, in which each agent's strategy is optimal given the strategies of the other agents. Correlated equilibrium (CE) is more general than NE in that it allows for dependencies among agents' strategies: a CE is a probability distribution over the agents' joint actions such that if a joint action is drawn from this distribution (presumably by a trusted third party), and each agent is told

V. Hakami (email: vhakami@aut.ac.ir) and M. Dehghan (email: dehghan@aut.ac.ir) are with the Department of Computer Engineering and Information Technology, Amirkabir University of Technology, PO Box 15875-4413, 424 Hafez Avenue, Tehran, Iran.

separately its own component, then it has no incentive to choose a different action, because, assuming that all others also obey, the suggested action is the best in expectation.

Stochastic games are particularly appealing since they capture both the strategic and stochastic aspects of a real-world scenario. Stochastic games with constraints [4] are even more interesting as they can also account for multiple objectives or for bounds on consumption of resources. In constrained stochastic games, the agents incur an additional cost at each stage which, similarly to the instantaneous reward, is a function of both the current state and the current action profile of the agents. The equilibrium policy should then be feasible under the agents' individual average/discounted constraints. Computational methods for equilibria in stochastic games have been actively pursued over the past decades. The majority of the schemes work in an offline fashion, i.e., for the case where Markovian dynamics (transition probabilities) are known a priori. Under this assumption, an extensive account on solution methods for stochastic games with special structures and with various reward criteria is given in [5]. Constrained stochastic games are typically approached via mathematical programming. See [4] and [6] for treatments of constrained games with jointly-controlled and independently-controlled state processes, respectively.

However, when Markovian dynamics are unknown, one may instead resort to learning-theoretic online solutions. It is within this perspective that stochastic games are also proposed as the standard framework for multi-agent reinforcement learning (MARL) [7]. Given the complexity of the strategy space in stochastic games, the solution concept sought in MARL algorithms is typically expressed in terms of *stationary* policies. The stationarity of a policy implies that it depends on the history of the game only through the current state. Traditionally, the MARL literature's prime interest has been directed towards learning stationary NE. Such equilibria have been shown to exist for both discounted and average reward stochastic games, with an extra ergodicity assumption on the transition structure of the latter [5]. These existence results also carry over to the constrained case, albeit with an additional strong Slater feasibility condition [4]. MARL algorithms for the computation of stationary NE in infinite-horizon general-sum stochastic games are primarily proposed for the unconstrained case. Depending on their informational assumptions, these algorithms can be divided into two broad classes: *joint action learning* (JAL) and *independent action learning* (IAL).

JAL algorithms constitute the early research on learning equilibria in stochastic games. These algorithms learn in the joint action space, and thus require that each agent observe the actions and possibly the rewards of the other agents. Prominent examples are Nash-$Q$ for discounted stochastic games [8], its variant, Nash-$R$ [9] for average reward games, and FF-$Q$ [10]. These are all multi-agent extensions of the celebrated *Q-learning* scheme of the MDP literature [11], with the distinction that they maintain $Q$-values for all possible *joint actions* at a given state. The major drawback associated with the current instances of JAL algorithms is that they all require repeated calls to an NE solver during the learning process, and this solver needs that the agents' $Q$-values be public information. Also, the convergence results are limited to a restricted class of games (e.g., common interest or zero-sum). In IAL algorithms, on the other hand, the agents only rely on their own past received rewards without knowing the actions or rewards of the other agents. IAL algorithms can thus operate in more informationally opaque scenarios, and unlike the case with JAL algorithms, their memory footprint is not exponential in the number of agents. A pioneer IAL algorithm is WOLF-PHC [12] which is only empirically evaluated and its convergence is not theoretically analyzed. The MG-ILA algorithm in [13] is based on an inter-connected learning automata abstraction and is only provably convergent in average reward games with pure NE policies. Finally, an ON-SGSP algorithm [14] has recently been proposed which is proved to be generally convergent to stationary NE policies in discounted games.

In this article, we depart from the NE-centric mainstream of MARL research, and instead address the problem of learning stationary CE in stochastic games [15]. The importance of CE arises from the fact that unlike NE, the concept of CE permits coordination between agents, and CE that are not NE can achieve higher rewards than NE, by avoiding positive probability mass on less desirable outcomes; in fact, CE payoffs in a one-shot game can fall outside the convex hull of all NE payoffs, and hence the former can make all players better off than the latter [3]. Within the context of normal-form games, the most efficient procedure for learning CE is the *no-regret* algorithm [16][17]. No-regret learning essentially requires that agents depart from their current play with probabilities that are proportional to measures of regret for not having used other strategies in the past. It is shown in [16] that once all the players' regrets approach zero, the joint empirical frequency of play converges with probability one to the set of CE of the game. A key property of no-regret learning is that it is an *uncoupled* update rule [18],[19]; i.e., each agent only needs to know its own reward function and to monitor the actions taken by the others to adjust its play probabilities.

When it comes to stochastic games, however, the literature on learning CE is very thin. The existence of stationary CE is implied by the existence of stationary NE in general-sum stochastic games. A direct proof is also given in [15] using a fixed point argument. As for the algorithms, CE-$Q$ [20] and $Q$nR [20][21] are the only MARL algorithms we know of that address the problem of learning stationary CE. Both algorithms belong to the JAL family, and use $Q$-learning to estimate the joint action values for each state of the game. Similarly to Nash-$Q$, CE-$Q$ relies on an equilibrium solver with access to all agents' $Q$-tables to update the CE policy in each iteration. Given the structural simplicity of CE w.r.t. to NE (convex polytope vs. fixed points), each call to a CE solver requires solving a set of linear inequalities, as opposed to an NE solver which has to deal with a nonlinear program. Yet again, the convergence results for CE-$Q$ are limited to zero-sum and common-interest games only. $Q$nR [20][21], on the other hand, is a fully decentralized algorithm. Realizing that a no-regret algorithm can serve as a natural learning backdrop for the agents to reach CE, $Q$nR eliminates the calls to a bulky equilibrium solver by interfacing $Q$-learning with no-regret-learning in a nested loop configuration. In the outer loop, the agents update their $Q$-values based on the empirical frequency of play that arises from the no-regret algorithm in the inner loop. The inner loop is in itself a virtual game governed by no-regret updates to which $Q$-values from the outer loop are fed as the agents' rewards. Hence, each iteration of the outer loop should essentially await the asymptotic convergence of the inner loop to zero-regret play probabilities. $Q$nR's main advantage is that it works without the luxury of a CE solver, and thus the agents' $Q$-tables remain private. In fact, owing to the uncoupledness of the no-regret algorithm, all an agent needs to observe is its opponents' play at each stage. Hence, $Q$nR itself can also be regarded as an uncoupled learning rule. The convergence of the algorithm, however, has not been analyzed in [21]. Moreover, it is challenging in practice to synchronize the agents for a virtual game in between two actual plays. Finally, the $Q$nR's nested loop configuration also makes it difficult to extend the algorithm to a constrained stochastic game setup.

*B. Contributions and Outline*

In this paper, we take the first step towards re-vitalizing interest in CE-centric MARL research by revamping $Q$nR in two ways: i) removing its virtual game interlude, and ii) extending it to also handle constrained games. In particular, we make the following contributions:

- Realizing that no-regret and $Q$-learning are both variants of stochastic approximation algorithms [17],[22], we exploit the multi-timescale extension of the theory of stochastic approximation to operate $Q$nR's inner and outer loops concurrently with two different step-size schedules. More specifically, we recast $Q$nR as a single-loop algorithm with no-regret learning moving on an effectively faster timescale than $Q$-learning. This way, we remove the virtual game interlude, while still preserving $Q$nR's main spirit: no-regret learning sees current $Q$-values as quasi-static, while $Q$-learning sees the estimated CE as essentially equilibrated.

- $Q$nR's recast as a stochastic approximation also makes it readily extensible to constrained setups. To show this, we first exploit the methodology in [15] to view the dynamics of the constrained game through the prism of a single agent. This is done by having each agent assume all the others adhere to the policy of an imaginary correlation device so that the environment reduces to a constrained MDP (CMDP) in its eyes. Using standard Lagrange duality [23] and the one-shot deviation principle of MDPs [24] we argue how the realization of CE in stochastic game amounts to simultaneous primal maximization in all agents' CMDPs. With this understanding, we may view the coupled iterates on joint policy and $Q$-values as primal ascent in individual agents' CMDPs which should then be augmented by a dual descent in Lagrange multiplier space. With $Q$nR's recast as a stochastic approximation, this augmentation can be done as easily as running stochastic sub-gradient descent on a slower third timescale. We refer to the overall algorithm as *constrained no-regret $Q$-learning* (CNR$Q$).

- Given the set-valued update increments of no-regret learning and the asynchronous nature of $Q$-learning iterations, CNR$Q$ would essentially correspond to a three-timescale asynchronous stochastic approximation with set-valued update increments. We give rigorous convergence results with differential inclusion arguments which draw on recent extensions of the theory of stochastic approximation to the case of asynchronous recursive inclusions with set-valued mean fields. The proof framework is due to Perkins and Leslie [25] who come up with conditions under which the asynchronicity of the process can be incorporated into the mean field to yield convergence results similar to those of an equivalent synchronous process. We verify that CNR$Q$ in fact satisfies these conditions and thus its asymptotic analysis can be facilitated via the arguments in [25].

- Finally, we present an example constrained stochastic game setup from the wireless networking domain. We use this example as a test bed to evaluate CNR$Q$'s performance and convergence behavior.

The rest of the paper is organized as follows: In II, we express the formalism of constrained general-sum stochastic games, with emphasis on both individual agent-level and system-wide control problems. In III, we present the machinery for learning stationary CE. To this end, we remark on the connection of both $Q$- and no-regret learning with the theory of stochastic approximation, and highlight the main idea in $Q$nR-learning, which paves way for the description of our CNR$Q$ algorithm. In IV, we establish CNR$Q$'s convergence. Finally, in V, we present numerical results for application of CNR$Q$-learning to an exemplary case from wireless networks. We conclude the paper in VI.

## II. Constrained General-Sum Stochastic Game

In this section, we begin with some notation and terminology that are associated with the definition of a constrained general-sum stochastic game. We then continue by formalizing the decision problem faced by each individual agent in II.A, and the social-level control problem in II.B which leads to the definition of a stationary correlated equilibrium. Finally, in II.C, we give an example embodiment of the game specification which serves both as a motivation for our algorithm in Section III and as a test bench to present numerical experiments in Section V.

A discrete-time, constrained stochastic game is denoted by a quintuple $\Gamma = \langle \mathcal{K}, \boldsymbol{A}, \mathcal{S}, (u_k(.))_{k \in \mathcal{K}}, (c_k(.))_{k \in \mathcal{K}} \rangle$. The elements constituting $\Gamma$ are defined as follows (see [4] for similar specifications):

- **Agents.** The agents participating in the game are indexed by the set $\mathcal{K} = \{1, 2, ..., K\}$, in which $K = |\mathcal{K}|$ (i.e., the cardinality of the set $\mathcal{K}$).

- **Actions.** We use $a_k^n \in A_k$ to denote the control action of the $k$-th agent at time $n=0,1,2,\ldots$. Let $\boldsymbol{a}^n = (a_1^n, .., a_K^n) \in \boldsymbol{A}$ denote the composition of the actions from all the agents at time $n$, where $\boldsymbol{A}$ is their joint action space. Also, denote by $\boldsymbol{a}_{-k}^n = (a_k^n)_{k \in \mathcal{K}, k \neq k}$ the action profile of agent $k$'s opponents at time $n$.

- **States.** The stochastic system state is modeled as a discrete time Markov decision chain (MDC). We use the random variable $s^n \in \mathcal{S} = \{1, 2, ..., S\}$ to indicate the system state at time $n$. We denote by $\mathcal{P}_{s\boldsymbol{a}\acute{s}}$ the transition probability between states $s$ and $\acute{s}$ under the joint action $\boldsymbol{a} \in \boldsymbol{A}$.

- **Instantaneous Utilities.** The utility $u_k^n$ accrued by each agent $k$ choosing an action $a_k^n$ at time $n$ can generally be expressed by a function $u_k: \mathcal{S} \times \boldsymbol{A} \to e$ of both the system state $s^n$ and the action profile $(a_k^n, \boldsymbol{a}_{-k}^n)$. $e$ denotes a compact interval in $\mathbb{R}$.

- **Instantaneous Constraints.** The immediate cost constraint $c_k^n$ incurred by each agent $k$ from choosing the control action $a_k^n$ at time $n$ is specified by a function $c_k: \mathcal{S} \times \boldsymbol{A} \to d$ of both the system state $s^n$ and the action profile $(a_k^n, \boldsymbol{a}_{-k}^n)$. $d$ denotes a compact interval in $\mathbb{R}$. We specify later that the cost constraints $c_k^n$ are involved in a long-term discounted constraint to be satisfied by the $k$-th agent.

- **Stationary Randomized Joint Policies.** Since we are interested in the set of CE of the game, it is easier to abstractly assume that there is a trusted third party (a referee or a correlation device in game-theoretic parlance) in the environment which is in charge of issuing recommendations to the agents at each stage of the game. Let $\boldsymbol{\pi}_s(.)$ be the policy used by the referee to sample joint plays at state $s$. $\boldsymbol{\pi}_s(.)$ is defined to be stationary in that it is a randomization over the joint action space $\boldsymbol{A}$ given only the current state $s$ and is independent of the history of the game. Each entry $\boldsymbol{\pi}_s(a_k, \boldsymbol{a}_{-k})$ represents the joint probability of taking action $a_k \in A_k$ by agent $k$ and action profile $\boldsymbol{a}_{-k} \in \boldsymbol{A}_{-k}$ by others at state $s$. We denote the entire set of the referee's joint policies over all states by $\Pi \triangleq \Delta(\boldsymbol{A})^{|\mathcal{S}|}$; i.e., $\boldsymbol{\pi}_s(.) \in \Delta(\boldsymbol{A})$.

The $n$-th stage of the game $\Gamma$ unfolds as follows: all agents and the referee observe the system state $s^n$; based on its policy $\boldsymbol{\pi}_{s^n}(.)$, the referee recommends an action $a_k^{ref,n}$ to each agent $k$. Given its recommendation, each $k$ chooses an action $a_k^n$, and the joint action $\boldsymbol{a}^n$ is played. Every agent $k$ accrues its payoff $u_k(s^n, \boldsymbol{a}^n)$, and incurs cost $c_k(s^n, \boldsymbol{a}^n)$. The play proceeds to stage $(n + 1)$ where $s^{n+1}$ is determined probabilistically according to $\mathcal{P}_{s^n \boldsymbol{a}^n s^{n+1}}$.

*A. Individual Agent's Control Problem*

Assume all other agents but $k$ play according to the referee's policy $\boldsymbol{\pi}$. Knowing $\boldsymbol{\pi}$ and given its recommended play $a_k^{ref}$ at state $s \in \mathcal{S}$, the agent $k$ can form a posteriori belief about the joint opponents' play $\boldsymbol{a}_{-k}$:

$$\pi_s(\boldsymbol{a}_{-k} \mid a_k^{ref}) = \frac{\pi_s(\boldsymbol{a}_{-k}, a_k^{ref})}{\sum_{\boldsymbol{b}_{-k} \in \boldsymbol{A}_{-k}} \pi_s(\boldsymbol{b}_{-k}, a_k^{ref})}. \quad (1)$$

Hence, from the point of view of the $k$-th agent, the environment reduces to a constrained MDP. Similarly to [15], in this MDP, we may break down the $n$-th stage of the play (from $n = 1,2, \dots$ onward) as: agent $k$ first observes the actions $\boldsymbol{a}_{-k}^{n-1}$ taken by its opponents in the previous round of $\Gamma$, perceives the payoff $u_k^{n-1}$ it has accrued during the $(n-1)$st stage together with its cost constraint $c_k^{n-1}$. It then observes the current state $s^n$, receives its advice $a_k^{ref,n}$ from the referee, and chooses an action $a_k^n$. We denote this CMDP by $\mathsf{M}_k = \langle \breve{A}_k, \breve{\mathcal{S}}_k, \breve{u}_k(.), \breve{c}_k(.) \rangle$ as follows:

- **Actions**. $\breve{A}_k = A_k$.

- **States**. $\breve{\mathcal{S}}_k = \{(\boldsymbol{a}_{-k}, s, a_k^{ref}) \in \boldsymbol{A}_{-k} \times \mathcal{S} \times A_k \mid \pi_s(a_k^{ref}) > 0\}$; i.e., we include in state $\breve{s}_k$ of agent $k$ from the stochastic game $\Gamma$, the previous actions of the other agents $\boldsymbol{a}_{-k}$, the current state $s$, and the referee's advice $a_k^{ref}$. The transition probabilities associated with this new state definition can be calculated as follows: Let $\breve{s}_k = (\boldsymbol{a}_{-k}, s, a_k^{ref})$ and $\acute{\breve{s}}_k = (\acute{\boldsymbol{a}}_{-k}, \acute{s}, \acute{a}_k^{ref})$. We have:

$$\breve{\mathcal{P}}_{\breve{s}_k a_k \acute{\breve{s}}_k} = \pi_s(\acute{\boldsymbol{a}}_{-k} | a_k^{ref}).\mathcal{P}_{s(a_k, \acute{\boldsymbol{a}}_{-k})\acute{s}}.\pi_{\acute{s}}(\acute{a}_k^{ref}). \quad (2)$$

- **Instantaneous Utility**. $\breve{u}_k(\breve{s}_k, a_k, \acute{\breve{s}}_k) = u_k(s, (a_k, \acute{\boldsymbol{a}}_{-k}))$.

- **Instantaneous Constraint**. $\breve{c}_k(\breve{s}_k, a_k, \acute{\breve{s}}_k) = c_k(s, (a_k, \acute{\boldsymbol{a}}_{-k}))$.

Let $\breve{\pi}_{k,\breve{s}_k}(.), \forall \breve{s}_k \in \breve{\mathcal{S}}_k$ denote agent $k$'s stationary policy, and consider a discount factor $\rho \in [0,1]$. Then, $k$'s discounted utility conditioned on initial state $\breve{s}_k \in \breve{\mathcal{S}}_k$ is defined as:

$$\breve{\mathcal{U}}_{k,\breve{s}_k}(\breve{\pi}_k) \overset{\text{def}}{=} \mathbb{E}\left[(1-\rho)\sum_{n=1}^{\infty}\rho^{n-1}\breve{u}_k(\breve{s}_k^n, a_k^n, \breve{s}_k^{n+1}) \big| \breve{s}_k^1 = \breve{s}_k\right], \quad (3)$$

where, the normalization factor $(1-\rho)$ ensures that the range of $\breve{\mathcal{U}}_k$ falls in the compact set $e^{|\breve{\mathcal{S}}_k|} \subset \mathbb{R}^{|\breve{\mathcal{S}}_k|}$. Now, for $\forall \breve{s}_k \in \breve{\mathcal{S}}_k$, the control problem faced by the $k$-th agent can be expressed as follows:

$$\max_{\breve{\pi}_k} \breve{\mathcal{U}}_{k,\breve{s}_k}(\breve{\pi}_k),$$
$$\textit{subject to the expected discounted cost constraint}: \quad (4)$$
$$\mathbb{E}\left[(1-\rho)\sum_{n=1}^{\infty}\rho^{n-1}\breve{c}_k(\breve{s}_k^n, a_k^n, \breve{s}_k^{n+1}) \big| \breve{s}_k^1 = \breve{s}_k\right] \leq \overline{\mathcal{D}}_k.$$

The constrained problem in (4) can be converted into an unconstrained form using standard Lagrangian approach [23],[26]. Let $\lambda_k \geq 0$ be a real number, called the Lagrange multiplier (LM). For agent $k$, define the instantaneous Lagrangian $\breve{\ell}_k : \mathbb{R}^+ \times \breve{\mathcal{S}}_k \times A_k \times \breve{\mathcal{S}}_k \to c$, where $c$ is a compact interval whose boundaries can be specified from $e, d$, and by ensuring that $\lambda_k$ is within an interval, say $[0, MAX] \subset \mathbb{R}^+$. The function $\breve{\ell}_k$ is defined as:

$$\breve{\ell}_k(\lambda_k, \breve{s}_k, a_k, \acute{\breve{s}}_k) \overset{\text{def}}{=} \breve{u}_k(\breve{s}_k, a_k, \acute{\breve{s}}_k) - \lambda_k(\breve{c}_k(\breve{s}_k, a_k, \acute{\breve{s}}_k) - \overline{\mathcal{D}}_k). \quad (5)$$

For $\forall \breve{s}_k \in \breve{\mathcal{S}}_k$, the expected total discounted Lagrangian associated with (5) is as follows:

$$\breve{\mathcal{L}}_{k,\breve{s}_k}^{\lambda_k}(\breve{\pi}_k) \stackrel{\text{def}}{=} \breve{\mathcal{U}}_{k,\breve{s}_k}(\breve{\pi}_k) - \lambda_k[\breve{\mathcal{C}}_{k,\breve{s}_k}(\breve{\pi}_k) - \overline{\mathcal{D}}_k] = \mathbb{E}\left[(1-\rho)\sum_{n=1}^{\infty}\rho^n\breve{\ell}_k(\lambda_k,\breve{s}_k^n,a_k^n,\breve{s}_k^{n+1})\Big|\breve{s}_k^1 = \breve{s}_k\right]. \quad (6)$$

The unconstrained counterpart to (4) is to determine the optimal pair $(\breve{\pi}_k^*, \lambda_k^*)$ such that the following saddle point optimality condition holds for $\forall \breve{s}_k \in \breve{\mathcal{S}}_k$ [26]:

$$\breve{\mathcal{L}}_{k,\breve{s}_k}^{\lambda_k^*}(\breve{\pi}_k) \leq \breve{\mathcal{L}}_{k,\breve{s}_k}^{\lambda_k^*}(\breve{\pi}_k^*) \leq \breve{\mathcal{L}}_{k,\breve{s}_k}^{\lambda_k}(\breve{\pi}_k^*). \quad (7)$$

With the (7) satisfied, $\breve{\mathcal{L}}_{k,\breve{s}_k}^{\lambda_k^*}(\breve{\pi}_k^*)$ is the optimal value of the problem (4), and it can be computed as [26]:

$$\breve{\mathcal{L}}_{k,\breve{s}_k}^{\lambda_k^*}(\breve{\pi}_k^*) = \min_{\lambda_k \geq 0} \max_{\breve{\pi}_k} \breve{\mathcal{L}}_{k,\breve{s}_k}^{\lambda_k}(\breve{\pi}_k), \forall \breve{s}_k \in \breve{\mathcal{S}}_k. \quad (8)$$

However, in the setup described by $\Gamma$, the maximization in (8) is solved concurrently by all agents, which undermines our simplifying single-agent abstraction. Next, we introduce a system-wide objective, which, when realized, amounts to $\breve{\mathcal{L}}_{k,\breve{s}_k}^{\lambda_k}(\breve{\pi}_k)$ being maximized simultaneously for all $k \in \mathcal{K}$.

### B. System-Wide Objective: Stationary Correlated Equilibria

Before giving a formal definition of stationary CE, we first express the long-term discounted Lagrangian of agent $k$ under the assumption that all agents (including $k$) follow the recommendations from a given referee's policy $\boldsymbol{\pi}$. Let $\boldsymbol{\lambda} = [\lambda_1, \ldots, \lambda_K]^T$ be a fixed vector of LMs for $\forall k \in \mathcal{K}$. Similarly to $\Pi$, define $\Pi^\lambda$ as the set of all stationary joint policies for the unconstrained version of the game $\Gamma$ with $\lambda_k$-parameterized individual Lagrangian utilities (denoted by $\Gamma^\lambda$ for easier reference). We have for $\forall s \in \mathcal{S}$:

$$\mathcal{L}_{k,s}^{\lambda_k}(\boldsymbol{\pi}) = \mathbb{E}\left[(1-\rho)\sum_{n=0}^{\infty}\rho^n\ell_k(\lambda_k,s^n,\boldsymbol{a}^n)|s^0 = s\right], \quad (9)$$

where, $\ell_k(\lambda_k,s,\boldsymbol{a}) \stackrel{\text{def}}{=} u_k(s,\boldsymbol{a}) - \lambda_k(c_k(s,\boldsymbol{a}) - \overline{\mathcal{D}}_k)$. It is well-known that $\mathcal{L}_{k,s}^{\lambda_k}$ has the following standard dynamic programming expansion (a.k.a. Bellman equations):

$$\mathcal{L}_{k,s}^{\lambda_k}\left(Q_k^{\lambda_k},\boldsymbol{\pi}\right) = \sum_{\boldsymbol{a}\in\mathcal{A}}\boldsymbol{\pi}_s(\boldsymbol{a}).Q_{k,(s,\boldsymbol{a})}^{\lambda_k}, \quad \forall s \in \mathcal{S} \quad (10)$$

where, $\mathcal{L}_{k,s}^{\lambda_k}$ is defined with an abuse of notation by making its dependence on $Q_k^{\lambda_k}$ explicit. $Q_k^{\lambda_k}: c^{|\mathcal{S}|} \rightarrow c^{|\mathcal{S}\times\mathcal{A}|}$ is a $|\mathcal{S}\times\mathcal{A}|$-dimensional $\lambda_k$-parameterized mapping whose $(s,\boldsymbol{a})$-th component evaluated at $\mathcal{L}_k^{\lambda_k}$ is defined as:

$$Q_{k,(s,\boldsymbol{a})}^{\lambda_k}\left(\mathcal{L}_k^{\lambda_k}\right) = (1-\rho).\mathbb{E}[\ell_k(\lambda_k,s,\boldsymbol{a})] + \rho\sum_{\acute{s}\in\mathcal{S}}\mathcal{P}_{s\boldsymbol{a}\acute{s}}\mathcal{L}_{k,\acute{s}}^{\lambda_k}\left(Q_k^{\lambda_k},\boldsymbol{\pi}\right). \quad (11)$$

Clearly, $Q_k^{\lambda_k}$ is an affine function of $\mathcal{L}_k^{\lambda_k}$, and the value function $\mathcal{L}_k^{\lambda_k}: c^{|\mathcal{S}\times\boldsymbol{A}|} \times \Pi^\lambda \rightarrow c^{|\mathcal{S}|}$ is a bilinear function of the policy $\boldsymbol{\pi}$ and action value function $Q_k^{\lambda_k}$. The dependence of $\mathcal{L}_k^{\lambda_k}$ on the policy as well as the inter-dependence of $\mathcal{L}_k^{\lambda_k}$ and $Q_k^{\lambda_k}$ is made explicit only on few occasions for emphasis. This dependence is otherwise suppressed to simplify notation. Now, we are ready to define $\Gamma^\lambda$'s set of stationary CE.

**Definition 1.** *The set $\mathsf{C}_{ce}^\lambda \subset \Pi^\lambda$ is called the set of stationary CE of the stochastic game $\Gamma^\lambda$ if under each $\boldsymbol{\pi}^{ce} \in \mathsf{C}_{ce}^\lambda$, it holds that for each agent $k$, for $\forall s \in \mathcal{S}$, for $\forall a_k^{ref} \in A_k$ with $\boldsymbol{\pi}_s^{ce}(a_k^{ref}) > 0$, and any alternative action $\acute{a}_k \in A_k$:*

$$\sum_{\boldsymbol{a}_{-k}\in\boldsymbol{A}_{-k}}\boldsymbol{\pi}_s^{ce}(\boldsymbol{a}_{-k}|a_k^{ref}).Q_{k,\left(s,\left(a_k^{ref},\boldsymbol{a}_{-k}\right)\right)}^{\lambda_k} \geq \sum_{\boldsymbol{a}_{-k}\in\boldsymbol{A}_{-k}}\boldsymbol{\pi}_s^{ce}(\boldsymbol{a}_{-k}|a_k^{ref}).Q_{k,(s,(\acute{a}_k,\boldsymbol{a}_{-k}))}^{\lambda_k}. \quad (12)$$

The inequality in (12) can be better understood if we intuitively consider $\Gamma^\lambda$ as a set of auxiliary normal-form games indexed by $s \in \mathcal{S}$ and with payoffs $Q^{\lambda_k}_{k,(s,\boldsymbol{a})}$ (e.g., see [1],[5]). By playing joint action $\boldsymbol{a}$ in the $s$-th auxiliary game, agent $k$'s payoff is the sum of its instantaneous payoff and the payoff it expects to gain from the next state onward, assuming joint policy $\boldsymbol{\pi}$. Now, $\boldsymbol{\pi}^{ce} \in \mathsf{C}^\lambda_{ce}$ if and only if it is simultaneously a CE for all auxiliary games $s \in \mathcal{S}$; i.e., if the referee draws its actions from $\boldsymbol{\pi}^{ce}$, $k$ realizes that every recommendation $a^{ref}_k$ it receives in each game $s \in \mathcal{S}$ is a best response to the estimated play of the other agents (assuming they all follow their recommendations). Now, we relate this collective notion with the agent-level objectives through the following theorem:

**Theorem I.** *For $\forall k \in \mathcal{K}$, $\forall \breve{s}_k \in \breve{\mathcal{S}}_k$, it holds that: $\breve{\mathcal{L}}^{\lambda_k}_{k,\breve{s}_k}(\breve{\pi}^*_k) = \mathcal{L}^{\lambda_k}_{k,s}\left(Q^{\lambda_k}_k, \boldsymbol{\pi}^{ce}\right)$.*

***Proof.*** As argued in ([15], Theorem 7), if all other agents but $k$ play according to the referee's policy $\boldsymbol{\pi}^{ce}$, then from the point of view of the $k$-th agent, the environment reduces to MDP $\mathsf{M}_k$, defined in II.$A$. By construction in [15], based on the one-shot deviation principle for MDPs [24], the referee's policy $\boldsymbol{\pi}^{ce}$ is a CE in the stochastic game if and only if its implementation in $\mathsf{M}_k$ is an optimal policy simultaneously for all $k \in \mathcal{K}$. It then follows that the expected discounted Lagrangian of all agents under the CE policy $\boldsymbol{\pi}^{ce}$ is equal to the expected discounted Lagrangian of the corresponding optimal policy in their MDPs. ∎

Now define Lagrange dual function $\mathcal{G}_k(\lambda_k) \overset{\text{def}}{=} \mathcal{L}^{\lambda_k}_k\left(Q^{\lambda_k}_k, \boldsymbol{\pi}^{ce}\right)$ as the solution of the primal problem $\max_{\breve{\pi}_k} \breve{\mathcal{L}}^{\lambda_k}_{k,\breve{s}_k}(\breve{\pi}_k)$ for $\forall \breve{s}_k \in \breve{\mathcal{S}}_k$ in (8). The optimal $\lambda^*_k$ can then be obtained by conducting dual descent on $\mathcal{G}_{k,s}(\lambda_k)$ for $\forall s \in \mathcal{S}$. In III, we present a distributed learning procedure to compute $\boldsymbol{\pi}^{ce}$ together with the optimal $\lambda^*_k$ for $\forall k \in \mathcal{K}$.

*C. Illustrative Example*

Before delving into the technicalities of learning a stationary CE, we give an illustrative example as typical real-world problems that can be modeled by the generic game described above. This example is a simplified yet an illustrative scenario from the domain of wireless networks which also provides a test bench to demonstrate both the convergence behavior as well as the efficacy of the algorithm discussed in Section III. The setting we consider is the resource control problem in hierarchical small-cell networks, more generally known as heterogeneous networks (HetNets) [27]. HetNets are wireless deployments where small cells (e.g., femto-cells) with lower signal power are positioned within the coverage area of a macro-cell primarily to multiply the capacity of this area. Traffic steering and load balancing are key aspects of HetNets as femto-cells can be installed in hotspots to offload much of the traffic from the macro layer. As envisioned in [28], HetNets will be integral to wireless deployments in near future. However, the coexistence of macro and small cell elements is not without ramifications. As argued in [28], HetNet topologies call for a new approach to run networks that is more complex, that requires a higher level of automation and more sophisticated resource control. Our examples in II.C.1 and II.C.2 entail uplink and downlink HetNet communication scenarios, respectively.

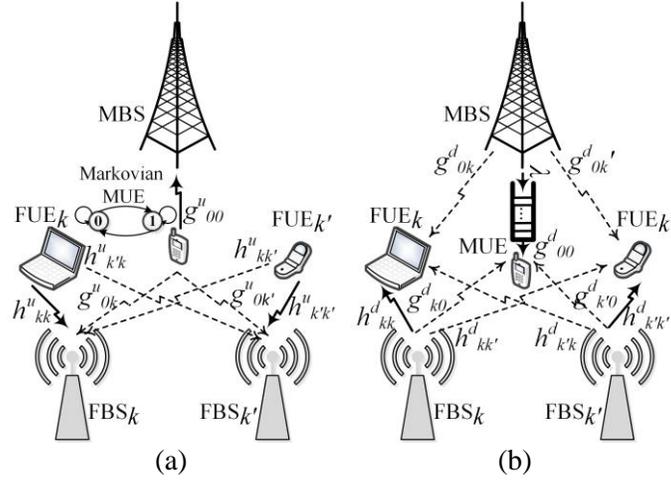

Fig. 1. Example constrained stochastic game setup: Decentralized resource control in two-tier small-cell networks. (a) uplink. (b) downlink.

### 1) *Spectrum Access Control in Uplink Communications*

Consider a two-tier CDMA femto-cell network (see Fig. 1(a)). It is assumed that the system consists of a single macro-cell base-station (MBS) receiving data from macro-user equipments (MUEs) in a region. Within this region, there are also $K$ co-channel femto-cells deployed by home or office users on the same frequency band (with bandwidth $W$) as the macro-cell. In each femto-cell, there is one femto base-station (FBS) receiving data from a number of femto user equipments (FUEs). For simplicity, only one active FUE is assumed in each cell. Let $h_{k\acute{k}}^{u}$ denote the gain of the link between FUE $k$ and FBS $\acute{k}$. Also, $No$ denotes the noise power on all channels. Each FUE seeks to maximize its own transmission rate which, by Shannon-Hartley's theorem (e.g., see [29]), depends on its received signal-to-interference plus noise ratio (SINR). We assume each FUE only gets to decide how aggressively (high/low in terms of power) it should transmit its signal; i.e., $a_k^{u,n} \in A_k = \{Low, High\}$. Due to the shared nature of the wireless channel, each FUE's perceived SINR depends not only on its own action but also on the actions of other FUEs. An FUE transmitting at a high power level, though may increase its own SINR, will interfere with the transmissions of the other FUEs, prompting them in turn to adopt a more aggressive behavior. Such a situation is undesirable since FUEs usually operate on limited batteries which require judicious consumption. In fact, the immediate cost $c_k^{u,n}$ incurred by each FUE $k$ from choosing action $a_k^{u,n}$ is its consumed power, i.e., $c_k^{u,n} = a_k^{u,n}$, with the restriction that the average power consumption over time should not exceed a pre-specified constraint $\bar{a}_k^u$. Moreover, given the two-tier structure of our setup, the activity of the MUEs is yet another source of interference, causing FUEs' signals to be further attenuated at their FBSs. Let $g_{0k}^u$ denote the channel gain between MUE and FBS $k$. MUE's interference activity over the shared channel is typically modeled as a time-homogenous discrete time Markov chain (DTMC) (e.g., see [30]). We use the binary random variable $s^n \in \mathcal{S} = \{0,1\}$ to indicate the macro activity at time $n$; i.e., $s^n = 1$ if the channel is occupied, in which case the interference power sensed at $k$-th FBS would be: $g_{0k}^u \cdot a_0$, where $a_0$ denotes the MUE's transmit power, and $g_{0k}^u$ denotes the gain of the link between MUE and FBS $k$. Also, $s^n = 0$, if the channel is idle. Hence, the uplink spectrum access control problem gives rise to a setting which is both strategic and stochastic. It is strategic since the FUEs' objectives are coupled due to mutual interference, and it is stochastic because FUEs' decisions have to be made under the effect of MUEs' Markovian dynamics. In this scenario, the utility $u_k^{u,n}$ accrued by each FUE $k$ at time

$n$ is its instantaneous Shannon rate:

$$u_k^{u,n} = u_k^u(s^n, a_k^{u,n}, \boldsymbol{a}_{-k}^{u,n}) = W.\log_2\left[1 + \frac{a_k^{u,n}.h_{kk}^u}{No + \mathbb{I}_{\{s^n\}}.g_{0k}^u.a_0^u + \sum_{k\in\mathcal{K}, k\neq k} a_k^{u,n}.h_{kk}^u}\right]. \quad (13)$$

Each FUE seeks a policy which maximizes its long-run rate utility subject to its power constraint. At the collective, social-level, it is desired to coordinate FUEs' decisions by striking a CE-based consensus. Our algorithm in III.B.3 allows FUEs to reach this consensus based only on their instantaneous rate and power consumption as feedbacks.

### 2) Power Control in Downlink Communications

In the same topology, consider the reverse scenario of downlink transmissions from MBS to its MUE and from FBSs down to their associated FUEs (Fig. 1(b)). We assume that MBS transmits at a constant power $a_0^d$, while each FBS chooses its power $a_k^{d,n}$ from a finite set of power levels. Let $h_{kk}^d$ denote the gain of the link between FBS $k$ and FUE $\acute{k}$; likewise, $\{g_{0k}^d\}_{k\in\mathcal{K}}$ (resp. $\{g_{k0}^d\}_{k\in\mathcal{K}}$) denotes MBS-FUE (resp., FBS-MUE) channel gains. Consistent with the common characterization of femto entities as best effort users, the traffic in FBS is assumed to be backlogged, while it is bursty and stochastic in MBS. Let $\mathcal{A}^n$ be the random number of packets arrived in the $n$-th timeslot to MBS's buffer whose capacity is capped by $N_B$ packets. The process $\{\mathcal{A}^n\}_{n\in\mathbb{N}}$ is assumed to be i.i.d. with general distribution $\mathbb{P}\{\mathcal{A}\}$ and mean $\lambda = \mathbb{E}[\mathcal{A}]$. By Shannon's law, MBS's achievable bit rate is given by (14) below:

$$r_0^n = W.\log_2\left[1 + \frac{a_0^d.g_{00}^d}{No + \sum_{k\in\mathcal{K}} a_k^{d,n}.g_{k0}^d}\right]. \quad (14)$$

Accordingly, the evolution of the system state (i.e., the buffer length in MBS) can be described by (15):

$$b_0^{n+1} = \min\left(\left(b_0^n - \frac{\tau.r_0^n}{L}\right)^+ + \mathcal{A}^n, N_B\right), \quad (15)$$

where, $\tau$ denotes the timeslot duration, $L$ is the packet length in bits, and $(.)^+$ stands for $\max(.,0)$. In this game, FBS agents are interested in maximizing their expected physical throughput (16) with the restriction that their interference to the macro layer be low enough so that the expected length of MBS's buffer remains below a certain threshold $\bar{b}_0$:

$$u_k^{d,n} = u_k^d(s^n, a_k^{d,n}, \boldsymbol{a}_{-k}^{d,n}) = W.\log_2\left[1 + \frac{a_k^{d,n}.h_{kk}^d}{No + g_{0k}^d.a_0^d + \sum_{k\in\mathcal{K}, k\neq k} a_k^{d,n}.h_{kk}^d}\right], \quad (16)$$

Again, it is desired that FBSs learn a stationary CE behavior by only receiving instantaneous feedbacks on their own rate $u_k^{d,n}$ and on MBS's buffer occupancy state $b_0^n$.

## III. Learning Stationary Correlated Equilibria

As with the case of MDPs, the fundamental update procedure for learning a policy can be derived from operationalizing Bellman equations in (10) and (11). For now, consider an unconstrained game, and imagine a centralized entity iteratively running the update equations below, for all $k \in \mathcal{K}$, for all $s \in \mathcal{S}$, and for all $\boldsymbol{a} \in \boldsymbol{A}(s)$:

$$\hat{V}_{k,s}^{n+1} := \sum_{\boldsymbol{a}\in\boldsymbol{A}(s)} \hat{\boldsymbol{\pi}}_s^n(\boldsymbol{a})\hat{Q}_{k,(s,\boldsymbol{a})}^n, \quad (17)$$

$$\hat{Q}_{k,(s,\boldsymbol{a})}^{n+1} := (1-\gamma).u_k(s,\boldsymbol{a}) + \gamma.\sum_{\acute{s}\in\mathcal{S}} \mathcal{P}_{s\boldsymbol{a}\acute{s}}\hat{V}_{k,\acute{s}}^{n+1}, \quad (18)$$

$$\hat{\boldsymbol{\pi}}_s^{n+1} \in \Pi^{ce}\left(\{\hat{Q}_{k,(s,.)}^{n+1}\}_{k\in\mathcal{K}}\right). \quad (19)$$

where, $\Pi^{ce}$ returns the set of all policies satisfying Definition 1. The first step to make this procedure more practical is to do without assuming that the matrix $\mathcal{P}$ of transition probabilities is known a priori. This is of particular interest in applications where the statistical knowledge regarding the processes underlying the system evolution is not available beforehand. In the context of our example from II.C, this corresponds to FUEs having to reach a consensus with no knowledge of the MUE's stochastic occupancy behavior, or FBSs adjusting their power levels without knowing the statistics of the packet arrival process in MBS (e.g., $\lambda$). The standard way to tackle the case of unknown $\mathcal{P}$ is to adopt an asynchronous variant of (18) with a decaying step size, better known as $Q$-learning. With this modification, the learning task would proceed by simulating a joint action, actually observing the next state, and run (18) for one state-action pair per learning iteration (see Table I for a procedure of this spirit).

The second step towards practice is decentralization so that each agent runs its local version of the above procedure. A naive decentralization, however, is subject to possible mis-coordination in the equilibrium selection step in (19). We discuss the known remedies for this issue as we review the existing CE learning algorithms in the sequel. Another issue has to do with the extent of knowledge an agent is assumed to have about its opponents. In fact, one can distinguish between coupled and uncoupled equilibrium learning procedures. In coupled learning, each agent needs to know the utilities of its opponents (e.g., the channel gains in (13) and (16)); uncoupled learning, however, is more practical since it can proceed without that luxury.

In this section, we first review the existing ideas for learning stationary CE. We begin by the coupled algorithm of CE-$Q$ [20], and highlight its shortcomings. Then, we turn to uncoupled $Q$nR learning [21], which prepares the ground for presentation of our CNR$Q$ algorithm, a provably convergent, constrained, and single-loop re-cast of $Q$nR.

*A. Existing Procedures for Learning Stationary CE in Stochastic Games*

CE-$Q$ [20] and $Q$nR [20][21] are the only algorithms we know of that address the problem of learning stationary CE. At each iteration, both CE-$Q$ and $Q$nR use $Q$-learning to update the $Q$-values based on the estimated CE policy for the next state. However, when it comes to update the CE policy itself, CE-$Q$ and $Q$nR differ significantly. In what follows, we briefly discuss the idea utilized in each algorithm for estimating CE and highlight their shortcomings.

*1) CE-Q Learning*

Much in the same way as the basic update rules in (17), (18), and (19), in CE-$Q$ [20], the estimate for the CE policy is obtained by solving the system of linear inequalities corresponding to the definition of CE (step 5 in Table I). Therefore, each agent $k$ is assumed to observe the rewards of all others and to maintain a model of their $Q$ tables. This requirement makes CE-$Q$ a coupled learning procedure. For ease of reference, we call this version of CE-$Q$ as semi-distributed. Also, as argued in [20], in the presence of multiple equilibria, *semi-distributed CE-Q* is subject to mis-coordination in the equilibrium selection step. This problem has been alleviated in [20] by introducing some equilibrium selection mechanisms. For instance, a *utilitarian* selector chooses an equilibrium which maximizes the sum of all agents' $Q$-values; however, except in very special games (e.g., zero-sum), CE-$Q$ in general needs that the play be centralized. We refer to this version of CE-$Q$ as *centralized CE-Q learning*.

*2) QnR Learning*

The $Q$nR algorithm in [21] eliminates the need for calls to an equilibrium solver. Instead, each agent relies on a no-regret learning algorithm to independently generate its own policy. Agents play according to their own policies, and compute their respective value functions based on the joint empirical distribution of play. This approach is theoretically sound since in the context of normal-form games, no-regret algorithms converge in empirical frequency to CE [16]. As we also rely on a no-regret procedure for our algorithm in III.B, we first give a brief account on the idea of no-regret learning, and then present $Q$nR's pseudo-code in Table II.

Consider a normal form game with payoff functions $(r_k(.))_{k \in \mathcal{K}}$. In no-regret learning [16], the agents reinforce the actions they regret not having played enough in the past. In particular, each agent $k$ has an implicit regret matrix $R_{k,\langle i,j \rangle}$ which maintains, for every pair of actions $i, j \in A_k$, the difference in average payoff if $k$ had taken action $j$ in the past every time it took action $i$; i.e.,

$$R_{k,\langle i,j \rangle}^n = \frac{1}{n} \sum_{\tau=1}^n [r_k(j, \boldsymbol{a}_{-k}^\tau) - r_k(i, \boldsymbol{a}_{-k}^\tau)].\, \mathbb{1}_{\{a_k^\tau = i\}}. \quad (20)$$

In $(n+1)$st round, given $a_k^n = i$, agent $k$ transitions to action $a_k^{n+1} = j$ with a probability $T_k^{n+1}(a^{n+1} = j | a_k^n = i)$ proportional to $R_{k,\langle i,j \rangle}^n$, and sticks to $i$ with $1 - \sum_{j \in A_k, j \neq i} T_k^{n+1}(a^{n+1} = j | a_k^n = i)$. In fact, the play probabilities $p_k^{n+1}$ for the next stage $(n+1)$ are obtained by solving the following balance equations:

$$p_k^{n+1}(i) \sum_{j \neq i} T_{k,\langle i,j \rangle}^{n+1} = \sum_{j \neq i} p_k^{n+1}(j). T_{k,\langle j,i \rangle}^{n+1}. \quad (21)$$

The learning proceeds by exploring choices and transitioning to actions which are conceived better according to the regret measure. Naturally, an agent's objective is to select a sequence of actions which guarantees to it no regret in the long run, no matter what the other agents do. Let $z^n(\boldsymbol{a})$ be the number of times the joint action profile $\boldsymbol{a}$ is actually played in the first $n$ periods, divided by $n$. In fact, $z^n(\boldsymbol{a})$ denotes the empirical distribution of play and is a probability distribution over $\boldsymbol{A}$. The no-regret learning of [16] has the property that when all agents' regret matrices approach to the non-positive orthant $\mathbb{R}_{-}^{|A_k \times A_k|}$, $z^n$ converges to the set of CE.

Now, in $Q$nR [21], in order to utilize the idea of no-regret learning in the context of stochastic games, each agent $k$ runs two nested control loops: $Q$-learning as the outer loop and multiple copies of the no-regret algorithm (one per state) as the inner loop (see Table II). At each outer loop iteration, the $s$-th copy of the no-regret algorithm of the inner loop starts afresh, fed by the current estimate of $Q$-values as $k$'s payoff function. The inner loop iterates until the agent's regret matrix $R_{k,s}^m$ converges to the non-positive orthant. Upon convergence of the inner loop, the outer loop begins its next iteration knowing that the joint empirical frequency of play for state $s$ will correspond to a CE of the game in $s$. $Q$nR learning goes on until all $Q$-table entries converge.

$Q$nR's advantage is that it works without requiring an equilibrium solver, and that the agents need not know their opponents' rewards to update their play probabilities. Hence, $Q$nR falls into the category of boundedly rational uncoupled learning dynamics [18][19]. $Q$nR's main disadvantage, however, is its nested loop structure. This not only makes it difficult to conduct a theoretical convergence analysis, but has some practical implications too: First, the virtual play in the inner loop, apart from being an interlude in the actual game, would require that the agents agree on a second iteration index during the learning process. The other limitation has to do with the extension of this paradigm to handle constrained problems, which leads to a third control loop and even more complications.

TABLE I
SKETCH OF THE CE-$Q$ ALGORITHM [20]

**Loop:**

**1.** Execute action $a_k$ in state $s$;

**2.** Observe joint agents' play $\boldsymbol{a}_{-k}$;

**3.** Observe own instantaneous reward $r_k(s, a_k, \boldsymbol{a}_{-k})$ as well as the reward $r_{\acute{k}}(.)$ for each agent $\acute{k} \in \mathcal{K} \backslash \{k\}$;

**4.** Observe next state $\acute{s}$;

**5.** Choose a CE policy $\boldsymbol{\pi}_{\acute{s}}^{CE} \in \Pi^{ce}\left(\left\{\hat{Q}_{k,(\acute{s},.)}\right\}_{k \in \mathcal{K}}\right)$ via solving the system of linear inequalities associated with CE;

**6. for each** agent $k \in \mathcal{K}$ **do**

    **6.1.** Update estimate for the value of the next state:
$$\hat{V}_{k,\acute{s}} := \sum_{\boldsymbol{b} \in \mathcal{A}} \boldsymbol{\pi}_{\acute{s}}^{CE}(\boldsymbol{b}) \hat{Q}_{k,(\acute{s},\boldsymbol{b})}^n;$$

    **6.2.** Update estimate for $Q$-value of current state-action pair:
$$\hat{Q}_{k,(s,\boldsymbol{a})} := \hat{Q}_{k,(s,\boldsymbol{a})} + \alpha\left[(1-\rho)\,r_k(s,\boldsymbol{a}) + \rho\hat{V}_{k,\acute{s}} - \hat{Q}_{k,(s,\boldsymbol{a})}\right];$$

**7.** Choose $\acute{a}_k$ (off-policy) and set $a_k := \acute{a}_k$;

**8.** Update $s := \acute{s}$; Go to 1;

TABLE II
SKETCH OF THE $Q$nR ALGORITHM [21]

**Outer Loop** (on-policy $Q$-learning):

**1.** Execute action $a_k$ and observe agents' play $\boldsymbol{a}_{-k}$;

**2.** Update joint empirical frequency of play $\hat{\boldsymbol{\pi}}_s(\boldsymbol{a})$;

**3.** Observe instantaneous reward $r_k(s, a_k, \boldsymbol{a}_{-k})$ and next state $\acute{s}$;

**4.** Update estimate for the value of the next state:
$$\hat{V}_{k,\acute{s}} := \sum_{\boldsymbol{a} \in \mathcal{A}} \hat{\boldsymbol{\pi}}_s(\boldsymbol{a}) . \hat{Q}_{k,(\acute{s},\boldsymbol{a})};$$
$$\hat{U}_k(s, a_k) := (1-\rho)r_k(s, a_k) + \rho\hat{V}_{k,\acute{s}};$$

**5.** Update estimate for $Q$-value of current state-action pair:
$$\hat{Q}_{k,(s,\boldsymbol{a})} := \hat{Q}_{k,(s,\boldsymbol{a})} + \alpha\left[\hat{P}_k(s, a_k) - \hat{Q}_{k,(s,\boldsymbol{a})}\right];$$

**6. Inner Loop** (no-regret learning):

    **for** $m = 0,1,2,\dots, M-1$ **do**

    **6.1.** Use (20) to update regret matrix $R_{k,(.,.),s}^{m+1}$ for state $s$ with $\hat{U}_k(s,.)$ as payoffs;

    **6.2.** Update the action transition probability matrix $T_{k,s}^{m+1}$;

    **6.3.** Use (21) to compute play probabilities $p_{k,s}^{m+1}$;

    **6.4.** Update average distribution of play:
$$\bar{p}_{k,s}^{m+1} := \frac{1}{m+1}\sum_{\tau=1}^{m+1} p_{k,s}^\tau;$$

**7.** Choose $\acute{a}_k$ by an $\varepsilon$-soft version of $\bar{p}_{k,s}^M(\acute{a}_k)$, and set $a_k := \acute{a}_k$;

**8.** Update $s := \acute{s}$; Go to 1;

## B. Proposed Algorithm

In this section, we present a stochastic approximation-based re-expression of $Q$nR which handles constrained games and, more importantly, is amenable to convergence analysis. As discussed in III.A.2, $Q$nR relies on the joint operation of no-regret and $Q$-learning working together in a nested loop configuration. Since both of these algorithms can be expressed in the form of a typical stochastic approximation [22][31],[32], we first very briefly remark on some general forms of stochastic approximation algorithms in III.B.1, and then highlight the connection of no-regret- and $Q$-leanring with relevant notions from the theory of stochastic approximation in III.B.2. Finally, in III.B.3, we give our version of things, referred to as CNR$Q$-learning.

### 1) Some General Forms of Stochastic Approximation Algorithms

Let $\mathcal{J} = \{1, \dots, |\mathcal{J}|\}$. A general stochastic recursive process has the following structure [32]:

$$x^{n+1} - x^n = \kappa(n)f(x^n, \vartheta^n), \quad (22)$$

where, $x^n \in \mathbb{R}^{|\mathcal{J}|}$, $f(x^n, \vartheta^n): \mathbb{R}^{|\mathcal{J}|} \times \mathbb{R}^{|\mathcal{J}|} \to \mathbb{R}^{|\mathcal{J}|}$, $\vartheta_n \in \mathbb{R}^{|\mathcal{J}|}$ is a random noise, and $\{\kappa(n)\}_{n \in \mathbb{N}}$ is a sequence of small, usually decreasing step-sizes. It is common to capture the noise effect as an additive term, by introducing: $F(x^n) = \mathbb{E}_\vartheta[f(x^n, \vartheta^n)]$, and $\mathcal{V}^{n+1} = f(x^n, \vartheta^n) - \mathbb{E}_\vartheta[f(x^n, \vartheta^n)]$, where, $F(x^n)$ is referred to as the mean field, and $\{\mathcal{V}^n\}_{n \in \mathbb{N}}$ is, by construction, a martingale difference sequence. In cases where the mean-field $F$ is a set-valued map (correspondence), we refer to the iteration above as a stochastic approximation with set-valued update increments or more concisely as a *stochastic recursive inclusion* [31]; i.e.,

$$x^{n+1} - x^n \in \kappa(n)[F(x^n) + \mathcal{V}^{n+1}].$$

Finally, let $2^\mathcal{J}$ be the power set of $\mathcal{J}$. If we denote by $\bar{J}^n \in 2^\mathcal{J}$ the components of the iterates $\{x^n\}_{n \in \mathbb{N}}$ updated at iteration $n$, we may use a counter $\chi^n(j) = \sum_{i=1}^{n} \mathbb{I}_{\{j \in j^i\}}$ to record how many times each component of $\{x^n\}_{n \in \mathbb{N}}$ have been updated until $n$. The following process, then, is called an *asynchronous stochastic approximation* [31] since it is no longer the case that all components of $x^n$ get updated simultaneously at time $n$:

$$x_j^{n+1} - x_j^n = \kappa(\chi^n(j)) . \mathbb{I}_{\{j \in j^n\}} [F_j(x^n) + \mathcal{V}_j^{n+1}],$$

*2) Q-Learning and No-Regret Learning as Stochastic Approximations*

Fix $\boldsymbol{\pi} \in \Pi$ as a stationary randomized policy over the joint action space $\boldsymbol{A}$. The function $Q_{k,(s,a)} \overset{\text{def}}{=} (1-\rho). \mathbb{E}[r_s(s, \boldsymbol{a})] + \rho \sum_{\acute{s} \in \mathcal{S}} \mathcal{P}_{s a \acute{s}} V_{k,\acute{s}}(\hat{Q}_k, \boldsymbol{\pi})$ is the expected long-term value of taking action $\boldsymbol{a}$ in state $s$, and following $\boldsymbol{\pi}$ thereafter. To learn this value without having to know about $\mathcal{P}$, one can use the $Q$-learning algorithm [11]. Due to [22], the exact form of the $Q$-learning update equation (e.g., step 5 in Table II) is as below:

$$\hat{Q}_{k,(s,a)}^{n+1} - \hat{Q}_{k,(s,a)}^n := \kappa(\chi^n(s, \boldsymbol{a})). \mathbb{I}_{\{(s,a)=(s^n, a^n)\}}. [F_{(s,a)}(\hat{Q}_k^n, \boldsymbol{\pi}) + \mathcal{V}_{k,(s,a)}^{n+1}], \quad (23)$$

with mean-field $F_{(s,a)}(\hat{Q}_k^n) = Q_{k,(s,a)} - \hat{Q}_{k,(s,a)}^n$, noise $\mathcal{V}_{(s,a)}^{n+1} = (1-\rho).r_k(s, \boldsymbol{a}) + \rho.V_{k,s^{n+1}}^n - Q_{k,(s,a)}$, and $V_{k,s}^n = \sum_{\boldsymbol{a}} \pi_s(\boldsymbol{a}).\hat{Q}_{k,(s,a)}^n$. Of particular note in is the use of asynchronous counters $\chi^n(s, \boldsymbol{a}) \overset{\text{def}}{=} \sum_{i=1}^{n} \mathbb{I}_{\{(s^i, a^i)=(s,a)\}}$. Such counters are needed as the time to visit each $(s, \boldsymbol{a})$ is random, and we might not have complete control over which component is to be updated next. Hence, by structure, $Q$-learning is an asynchronous stochastic approximation.

As for no-regret learning, it is noted in [17] that (20) can be cast as a moving average with step size $\kappa(n) = 1/n$:

$$R_k^{n+1} - R_k^n = \kappa(n). \left( [r_k(j, \boldsymbol{a}_{-k}^n) - r_k(i, \boldsymbol{a}_{-k}^n)]. \mathbb{I}_{\{a_k^n = i\}} - R_k^n \right), \quad (24)$$

Also, it has been shown in [17][17] that the above equation has the following correspondence $F(R_k^n)$ as its mean field:

$$F(R_k^n) = C_k(p_k^n \times \Delta(\boldsymbol{A}_{-k})) - R_k^n, \quad (25)$$

where, for $x \in \Delta(\boldsymbol{A})$, $C_k(x)$ is a $|A_k| \times |A_k|$ matrix with entries:

$$C_{k,(i,j)}(x) \overset{\text{def}}{=} \sum_{\boldsymbol{a} \in A: a_k = i} x(\boldsymbol{a}). [r_k(j, \boldsymbol{a}_{-k}) - r_k(i, \boldsymbol{a}_{-k})]. \quad (26)$$

Hence, the regret update procedure can be re-written as a stochastic recursive inclusion of the form:

$$R_k^{n+1} - R_k^n \in \kappa(n). \left( F(R_k^n) + \mathcal{V}_k^{n+1} \right), \quad (27)$$

where, the random noise term $\mathcal{V}_k^{n+1}$ is as below:

$$\mathcal{V}_k^{n+1} \in [r_k^n(j, \boldsymbol{a}_{-k}^n) - r^n(i, \boldsymbol{a}_{-k}^n)]. \mathbb{I}_{\{a_k^n = i\}} - C_k(p_k^n \times \Delta(\boldsymbol{A}_{-k})). \quad (28)$$

*3) CNRQ Learning*

In III.B.2, we remarked on the fact that both no-regret- and $Q$-learning are, by structure, special cases of stochastic approximation algorithms. In this section, we resort to the multi-timescale extension of standard stochastic approximation theory [32] to recast the nested loop structure of the $Q$nR algorithm as a single-loop two-timescale stochastic approximation. The idea is to have the $Q$-learning and no-regret iterations proceed simultaneously with different step-size schedules so that $Q$-table entries get updated on a slower effective timescale compared to the regret-matrix updates. Multi-timescale arguments of stochastic approximation [32] then guarantee that no-regret iterations see $Q$-learning as quasi-static while the latter sees the former as nearly equilibrated, thus mimicking the $Q$nR's nested loop configuration. Following the same methodology, we introduce an even slower third timescale for updating the Lagrange multipliers associated with the game's constraints. More specifically, we leverage on Theorem I and the saddle point property in (7) to cast the algorithm as a primal-dual scheme; i.e., given a fixed $\lambda_k$ for each agent $k$,

'primal' maximization reduces to computing, in a distributed fashion, a CE behavior $\boldsymbol{\pi}^{ce}$ (c.f., Definition 1) of the stochastic game $\Gamma^\lambda$. Also, given that $\mathcal{G}_k(\lambda_k) = \mathcal{L}_k^{\lambda_k}\left(Q_k^{\lambda_k}, \boldsymbol{\pi}^{ce}\right)$, the correct multiplier $\lambda_k^*$ can be learned by stochastic gradient descent in the 'dual' space, performed on the slowest timescale, so that it sees the 'primal' maximization as having essentially equilibrated. We refer to the overall algorithm as CNR$Q$-learning. Given the set-valued update increments of no-regret learning and the asynchronous nature of the $Q$-learning iterations, CNR$Q$ would essentially correspond to a three-timescale asynchronous stochastic recursive inclusion. We save the formalization of these ideas for Section IV, where we give a detailed convergence analysis. Here, we mainly establish notation and discuss the algorithm's workflow. Let the learning rates $\{\alpha(n)\}_{n\in\mathbb{N}}$, $\{\beta(n)\}_{n\in\mathbb{N}}$, and $\{\gamma(n)\}_{n\in\mathbb{N}}$ satisfy (A1) below:

$$\sum_n \alpha(n) = \sum_n \beta(n) = \sum_n \gamma(n) = \infty.$$
$$\sum_n (\alpha(n)^2 + \beta(n)^2 + \gamma(n)^2) < \infty. \qquad \text{(A1)}$$
$$\frac{\alpha(n)}{\gamma(n)}, \frac{\beta(n)}{\alpha(n)} \to 0 \quad as \ n \to \infty.$$

Also, for $\forall s \in \mathcal{S}$ and $\forall \boldsymbol{a} \in \boldsymbol{A}$, let $\phi^n(s)$ and $\upsilon^n(s, \boldsymbol{a})$ be two asynchronous counters: $\phi^n(s) := \sum_{i=1}^n \mathbb{I}_{\{s^i = s\}}$ and $\upsilon^n(s, \boldsymbol{a}) := \sum_{i=1}^n \mathbb{I}_{\{(s^i, \boldsymbol{a}^i)=(s, \boldsymbol{a})\}}$. We organize the CNR$Q$'s workflow into 9 steps, as listed in Table III:

- In step 0, each agent initializes the empirical frequency of joint play $\boldsymbol{\pi}_s^0(.)$, Lagrange multiplier $\lambda_k^0$, $Q$-table $\hat{Q}_{k,(s,.)}^0$, and state-dependent regret matrix $R_{k,(.,.),s}^0$. It then samples its action $a_k^0$ from a uniform distribution.

- In step 1, according to the observed joint opponents' play $\boldsymbol{a}_{-k}^n$, agent $k$ updates the empirical distribution $\boldsymbol{\pi}_s^n(.)$ on the fastest timescale. It would be convenient to express $\boldsymbol{\pi}_s^n(.)$ in closed-form as below:

$$\boldsymbol{\pi}_s^n(\boldsymbol{a}) = \sum_{\eta \le n} \gamma\left(\phi^{\eta-1}(s)\right) \left[\prod_{\zeta=\eta}^{n-1}\left(1 - \gamma\left(\phi^\zeta(s)\right)\right)\right] e_{\boldsymbol{a}^n}. \qquad (29)$$

- In step 2, agent $k$ calculates its instantaneous Lagrangian $\ell_k(\lambda_k^n, s^n, \boldsymbol{a}_{-k}^n)$ for its played action $a_k^n$, and observes the next state of the system $s^{n+1}$.

- Steps 3 and 4 update the $Q$-table $\hat{Q}_k^n$ using $Q$-learning on the moderate timescale. This step unfolds as follows: agent $k$ first computes its Lagrangian value function $\mathcal{L}_{k,s^{n+1}}^{\lambda_k}$ for the next state $s^{n+1}$ based on the empirical frequency of play $\boldsymbol{\pi}^{n+1}$ and current estimate $\hat{Q}_k^n$. It then updates its $Q$-table using both its instantaneous Lagrangian $\ell_k$ and its long-term Lagrangian $\mathcal{L}_k^{\lambda_k}$.

- In step 5, Lagrange multiplier $\lambda_k^n$ is updated based on the perceived cost $c_k(s^n, \boldsymbol{a}^n)$ and using stochastic (sub-)gradient descent on the slowest timescale.

- Step 6 is devoted to the state-dependent regret matrix update. The regret matrix $R_{k,(i,j),s}^n$ is conditional on $k$'s current play $a_k^n = i$, and is calculated as the $Q$-value differential between $i$ and every alternative action $j$. Similarly to step 1, this update equation runs on the fastest timescale. To make more explicit the dependency of $R_k^n$ on both $\hat{Q}_k^n$ and $\boldsymbol{\pi}^n$, one may use (29) to re-write $R_{k,(i,j),s}^n$ for $\forall i, j \in A_k$ as follows:

$$R_{k,\langle i,j\rangle,s}^n = \sum_{\eta \le n:a_k^\eta=i} \gamma\left(\phi^{\eta-1}(s)\right)\left[\prod_{\zeta=\eta}^{n-1}\left(1-\gamma\left(\phi^\zeta(s)\right)\right)\right]\cdot\left(\hat{Q}_{k,(s,j,\boldsymbol{a}_{-k}^n)}^\eta - \hat{Q}_{k,(s,i,\boldsymbol{a}_{-k}^n)}^\eta\right)$$

$$= \sum_{\boldsymbol{a}\in A:a_k=i} \boldsymbol{\pi}_s^n(\boldsymbol{a})\cdot\left(\hat{Q}_{k,(s,j,\boldsymbol{a}_{-k}^n)}^n - \hat{Q}_{k,(s,i,\boldsymbol{a}_{-k}^n)}^n\right). \tag{30}$$

- **Step 7** uses the updated regret-values to compute the action transition probabilities $T_{k,s}^{n+1}\left(a_k^n = j \middle| a_k^n = i\right)$ from the current action $i$ to every alternative action $j$. $T_k^{n+1}$ is proportional to the positive part of the regret measure; i.e., $\max(R_{k,s}^{n+1}, 0)$. However, to ensure the smoothness of these transitions, we use a function $\Upsilon(.)$ as a smooth version of $\max(.,0)$, defined as: $\Upsilon(x) \triangleq \begin{cases} x, & x > 0 \\ 0, & x < 0 \end{cases}$ for any $\delta > 0$ and $x \notin \delta - neighborhood(0)$.

- Finally, in **step 8**, the action for the next stage $(n+1)$ is sampled from $p_{k,s^{n+1}}^{n+1}$ which is an $\varepsilon$-soft version of the regret-based strategy $\hat{p}_{k,s^{n+1}}^{n+1}$ with $\varepsilon$ being the exploration factor. $\hat{p}_k^{n+1}$ is an invariant measure for the stochastic transition matrix $T_k^{n+1}$. Therefore, $p_k^{n+1}$ can be viewed as the invariant measure for the $\varepsilon$-trembled version of $T_k^{n+1}$ denoted by $\bar{T}_k^{n+1}$, and can be obtained by solving the following balance equations for $\forall s \in \mathcal{S}, \forall a \in A_k$:

$$p_{k,s}^{n+1}(a)\sum_{\acute{a}\in A_k-\{a\}} \bar{T}_{k,\langle a,\acute{a}\rangle,s}^{n+1} = \sum_{\acute{a}\in A_k-\{a\}} p_{k,s}^{n+1}(\acute{a})\cdot\bar{T}_{k,\langle\acute{a},a\rangle,s}^{n+1}, \tag{31}$$

$$\text{where,} \quad \bar{T}_{k,\langle i,j\rangle,s}^{n+1} := (1-\varepsilon)\frac{\Upsilon\left(R_{k,\langle i,j\rangle,s}^{n+1}\right)}{\mu} + \frac{\varepsilon}{|A_k|}, \quad \forall i,j\in A_k. \tag{32}$$

TABLE III
CONSTRAINED NO-REGRET $Q$-LEARNING (CNR$Q$-LEARNING)

---

**0) Initialization:** for $\forall s \in \mathcal{S}$, set: $\boldsymbol{\pi}_s^0(.) = 0$; $\hat{Q}_{k,(s,.)}^0 = 0$; $\lambda_k^0 = 0$; $R_{k,\langle.,.\rangle,s}^0 = 0$; $a_k^0 \sim \frac{1}{|A_k|}$;

for $n = 0,1,2,...$ repeat the following steps:

1) **Observe the joint opponents' play** $\boldsymbol{a}_{-k}^n$ and update the empirical frequency of $\forall \boldsymbol{a} \in A$ for $\forall s \in \mathcal{S}$:
$\boldsymbol{\pi}_s^{n+1}(\boldsymbol{a}) := \boldsymbol{\pi}_s^n(\boldsymbol{a}) + \gamma(\phi^n(s)).\mathbb{I}_{\{s=s^n\}}.\left(\boldsymbol{e}_{a^n} - \boldsymbol{\pi}_s^n(\boldsymbol{a})\right);$ // $\boldsymbol{e}_a$ is the unit vector in $\Delta(A)$ w.r.t. $\boldsymbol{a} \in A$

2) **Observe utility** $u_k(s^n, \boldsymbol{a}^n)$, **cost** $c_k(s^n, \boldsymbol{a}^n)$, **calculate Lagrangian** $\ell_k(\lambda_k^n, s^n, \boldsymbol{a}^n)$, **observe next state** $s^{n+1}$.

3) **Calculate the long-term Lagrangian:** $\mathcal{L}_{k,s^{n+1}}^{\lambda_k} := \sum_{\acute{a}\in\mathcal{A}} \boldsymbol{\pi}_{s^{n+1}}^{n+1}(\acute{a}).\hat{Q}_{k,(s^{n+1},\acute{a})}^n$;

4) **Update the Lagrangian $Q$-table for $\forall s \in \mathcal{S}, \forall \boldsymbol{a} \in A$ using $Q$-learning:**
$\hat{Q}_{k,(s,\boldsymbol{a})}^{n+1} := \hat{Q}_{k,(s,\boldsymbol{a})}^n + \alpha(v^n(s,\boldsymbol{a})).\mathbb{I}_{\{(s,\boldsymbol{a})=(s^n,\boldsymbol{a}^n)\}}.\left[(1-\rho).\ell_k(\lambda_k^n,s,\boldsymbol{a}) + \rho.\mathcal{L}_{k,s^{n+1}}^{\lambda_k} - \hat{Q}_{k,(s,\boldsymbol{a})}^n\right];$

5) **Update Lagrange multiplier using stochastic (sub-)gradient descent:**
$\lambda_k^{n+1} := [\lambda_k^n + \beta(n).(c_k(s^n,\boldsymbol{a}^n) - \bar{\mathcal{D}}_k)]^+;$   // $[.]^+$ denotes projection onto $[0, MAX]$

6) **Update the $|A_k| \times |A_k|$ state-dependent regret matrix $R_k^n$ for $\forall i,j \in A_k$ and $\forall s \in \mathcal{S}$:**
$R_{k,\langle i,j\rangle,s}^{n+1} := R_{k,\langle i,j\rangle,s}^n + \gamma(\phi^n(s)).\mathbb{I}_{\{s=s^n\}}.\left[\left(\hat{Q}_{k,(s,j,\boldsymbol{a}_{-k}^n)}^{n+1} - \hat{Q}_{k,(s,i,\boldsymbol{a}_{-k}^n)}^{n+1}\right).\mathbb{I}_{\{a_k^n=i\}} - R_{k,\langle i,j\rangle,s}^n\right];$

7) **Update the action transition probability for $\forall s \in \mathcal{S}$ and $\forall j \in A_k$:**
$$T_{k,s}^{n+1}(a_k^{n+1}=j|a_k^n=i):=\begin{cases}\dfrac{\Upsilon\left(R_{k,\langle i,j\rangle,s}^{n+1}\right)}{\mu}, & j \neq i\\ 1-\displaystyle\sum_{l\in A_k,l\neq j} T_{k,s}^{n+1}(a_k^{n+1}=l|a_k^n=i), & j = i\end{cases};$$
where, $\mu$ is the inertia constant and is large enough to ensure that $T_{k,s}^{n+1}(a_k^{n+1}=i|a_k^n=i) > 0$ for $\forall i \in A_k$.

8) **Action selection:** Choose action $a_k^{n+1}$ in state $s^{n+1}$ according to the following distribution:
$p_{k,s^{n+1}}^{n+1} := (1-\varepsilon).\hat{p}_{k,s^{n+1}}^{n+1} + \dfrac{\varepsilon}{|A_k|}.\mathbf{1}_{|A_k|},$
where, $\hat{p}_k^{n+1}$ is an invariant measure for $T_k^{n+1}$, $\mathbf{1}_{|A_k|}$ is a $|A_k| \times 1$ vector of 1's, and $0 < \varepsilon \ll 1$ is a small tremble.

---

# IV. Convergence Analysis

CNR$Q$-learning is essentially a three-timescale asynchronous stochastic recursive inclusion. To establish CNR$Q$'s convergence, we exploit the recent results by Perkins and Leslie [25] which facilitate the asymptotic analysis in cases such as ours where the update patterns involved are both asynchronous and set-valued. The proof framework we use is called *asynchronous stochastic approximation with differential inclusions*. The results given in [25] already account for two-timescale setups as well. Also, since in general, the ideas underlying the multi-timescale arguments carry over when the number of timescales is more than two [32], using the two-timescale analysis in [25], we first analyze the coupled recursions of no-regret- and $Q$-learning by freezing $\lambda_k^n \approx \lambda_k$; in fact, CNR$Q$'s iterates can be interpreted as a primal-dual scheme, with $(\hat{Q}_k^n, \boldsymbol{\pi}^n)$ getting updated by primal iterations and $\lambda_k^n$ by dual iterations. Now, in view of $\beta(n) = o(\alpha(n))$, the dual minimization is carried out at a slower timescale so that it sees the primal maximization as equilibrated while the latter sees the former as quasi-static. The analysis of the pair $(\hat{Q}_k^n, \boldsymbol{\pi}^n)$ can be conducted by invoking the results of ([25], Section 4). Once the almost sure convergence of the primal iterates is established, we have $d(\boldsymbol{\pi}^n, \mathsf{C}_{ce}^{\lambda}) \to 0$, $\hat{Q}_k^n - Q_k^{\lambda_k} \xrightarrow{n \uparrow \infty} 0$, and thus $\mathcal{L}_k^{\lambda_k}(\hat{Q}_k^n, \boldsymbol{\pi}^n) \to \mathcal{G}_k(\lambda_k)$. Then, using results from constrained reinforcement learning (e.g., see [33]), we can prove that the dual iterates $\lambda_k^n$ also converge to $\lambda_k^*$. We organize our convergence analysis into three parts: first, we extract the mean-field and noise components associated with the primal iterates $(\hat{Q}_k^n, \boldsymbol{\pi}^n)$ in IV.$A$. Next, in IV.$B$, we verify the conditions which should be satisfied by these components so that the results in [25] become applicable to our case. Finally, in IV.$C$, we come up with differential inclusion arguments to establish the convergence of CNR$Q$ along the lines of ([25], Section 4).

## A. Identifying the Mean-Field and Noise Components

According to Table I, the estimates $\{\boldsymbol{\pi}^n\}_{n \in \mathbb{N}}$ and $\{\hat{Q}_k^n\}_{n \in \mathbb{N}}$ are given iteratively by the following coupled process:

$$\hat{Q}_{k,(s,\boldsymbol{a})}^{n+1} - \hat{Q}_{k,(s,\boldsymbol{a})}^n = \alpha\big(v^n(s,\boldsymbol{a})\big). \mathbb{I}_{\{(s,\boldsymbol{a})=(s^n,\boldsymbol{a}^n)\}}. \big[ F_{(s,\boldsymbol{a})}\big(\hat{Q}_k^n, \boldsymbol{\pi}^n\big) + V_{k,(s,\boldsymbol{a})}^{n+1} \big], \quad (33)$$

$$\boldsymbol{\pi}_s^{n+1} - \boldsymbol{\pi}_s^n \in \gamma\big(\phi^n(s)\big). \mathbb{I}_{\{s=s^n\}}. \big[ G_s\big(\hat{Q}_k^n, \boldsymbol{\pi}^n\big) + U_{k,s}^{n+1} \big], \quad (34)$$

where,

$$F_{(s,\boldsymbol{a})}\big(\hat{Q}_k^n, \boldsymbol{\pi}^n\big) = \mathrm{H}_{(s,\boldsymbol{a})}^{\lambda_k}\big(\hat{Q}_k^n, \boldsymbol{\pi}^n\big) - \hat{Q}_{k,(s,\boldsymbol{a})}^n, \quad (35)$$

$$G_s\big(\hat{Q}_k^n, \boldsymbol{\pi}^n\big) = \Psi_s\big(\hat{Q}_k^n, \boldsymbol{\pi}^n\big) - \boldsymbol{\pi}_s^n. \quad (36)$$

The mapping $\mathrm{H}_{(s,\boldsymbol{a})}^{\lambda_k}(.,.)$ in (35) is defined for general $\hat{Q}_k \in c^{|\mathcal{S} \times A|}$ and $\boldsymbol{\pi} \in \Pi^{\lambda}$ as:

$$\mathrm{H}_{(s,\boldsymbol{a})}^{\lambda_k}\big(\hat{Q}_k, \boldsymbol{\pi}\big) = (1-\rho). \mathbb{E}[\ell_k(\lambda_k, s, \boldsymbol{a})] + \rho \sum_{\acute{s} \in \mathcal{S}} \mathcal{P}_{s\boldsymbol{a}\acute{s}} \mathcal{L}_{k,\acute{s}}^{\lambda_k}\big(\hat{Q}_k, \boldsymbol{\pi}\big). \quad (37)$$

$\Psi_s$ in (36) is a correspondence evaluated at a given $(\hat{Q}_k, \boldsymbol{\pi})$ as:

$$\Psi_s\big(\hat{Q}_k, \boldsymbol{\pi}\big) \triangleq \{p_{k,s} \times \Delta(\boldsymbol{A}_{-k}) \,|\, p_{k,s} \text{ satisfies } (14)\}. \quad (38)$$

Note that in view of (32) and (30), $\Psi_s$ is dependent on both $\hat{Q}_k$ and $\boldsymbol{\pi}$.

To specify the stochastic components $\{V_k^n\}_{n \in \mathbb{N}}, \{U_k^n\}_{n \in \mathbb{N}}$, let $\bar{H} = \{\big((s,\boldsymbol{a}),s\big); s \in \mathcal{S}, a \in \boldsymbol{A}\}$, with $\bar{H}^n \in \bar{H}$ being the updated component across $\bar{H}$ at iteration $n$; also, let $z^n = \big(\hat{Q}_k^n, \boldsymbol{\pi}^n\big)$. Define $\mathcal{F}_n$ as the $\sigma$-algebra containing all the information up until the end of the $n$-th iteration; i.e., $\mathcal{F}_n \triangleq \sigma\big(\{\bar{H}^m\}_m, \{z^m\}_m, \{v^m(s,\boldsymbol{a})\}_{(s,\boldsymbol{a}),m}, \{\phi^m(s)\}_{s,m}\big)$; $\forall m \leq$

$n$, $s \in \mathcal{S}$, $(s, \boldsymbol{a}) \in \mathcal{S} \times \boldsymbol{A}$. Then, $\{V_k^n\}_{n \in \mathbb{N}}$, $\{U_k^n\}_{n \in \mathbb{N}}$ are, by construction, $\mathcal{F}_n$-adapted martingale difference processes defined on $\mathbb{R}^{|\mathcal{S} \times \boldsymbol{A}|}$ and $\mathbb{R}^{|\mathcal{S}|}$ resp. as follows:

$$V_{k,(s,\boldsymbol{a})}^{n+1} = (1 - \rho) . \ell_k(\lambda_k^n, s, \boldsymbol{a}) + \rho . \mathcal{L}_{k,s^{n+1}}^{\lambda_k} - \mathrm{H}_{(s,\boldsymbol{a})}^{\lambda_k}(\hat{Q}_k^n, \boldsymbol{\pi}^n), \quad (39)$$

$$U_{k,s}^{n+1} \in e_{\boldsymbol{a}^n} - \Psi_s(\hat{Q}_k^n, \boldsymbol{\pi}^n). \quad (40)$$

### B. Verifying Technical Assumptions

In this section, we verify the conditions required by [25] on the mean-field and noise components of the $(\hat{Q}_k^n, \boldsymbol{\pi}^n)$ iterates. We do this by presenting a sequence of lemmas (I to IV) corresponding resp. to ([25], Assumptions: (B1), (B3), (B4), and (B5)). Assumption (B2) in [25] is already satisfied by our assumption (A1) on step-sizes in III.B.3. Please refer to Appendix A in the supplementary materials of this paper for proofs of lemmas I to IV.

**Lemma I.** *For compact sets, $C \subset \mathbb{R}^{|\mathcal{S} \times \boldsymbol{A}|}$, $D \subset \mathbb{R}^{|\mathcal{S}|}$, $\hat{Q}_k^n \in C$, $\boldsymbol{\pi}^n \in D$ for all $k$ and $n$.*

**Lemma II.** *It holds that:*

(a) *$G(.,.) : c^{|\mathcal{S} \times \boldsymbol{A}|} \times \Pi^\lambda \to \Pi^\lambda$ is a Marchaud map [34]; i.e., (i) the graph and domain of $G$ are non-empty and closed, (ii) the values $G(\hat{Q}_k, \boldsymbol{\pi})$ are convex, and (iii) the growth of $G$ is linear.*

(b) *$F(.,.) : c^{|\mathcal{S} \times \boldsymbol{A}|} \times \Pi^\lambda \to c^{|\mathcal{S} \times \boldsymbol{A}|}$ is upper semi-continuous, and for all $\boldsymbol{\pi} \in \Pi^\lambda$, $F(., \boldsymbol{\pi}) : c^{|\mathcal{S} \times \boldsymbol{A}|} \to c^{|\mathcal{S} \times \boldsymbol{A}|}$ is a Marchaud map.*

**Lemma III.** *Consider $\mathcal{H}_n, \mathcal{H}_{n+1} \in \bar{H}$, then:*

(a) *$\mathbb{P}(\bar{H}_{n+1} = \mathcal{H}_{n+1} \mid \mathcal{F}_n) = \mathcal{Q}_{(\mathcal{H}_n \mathcal{H}_{n+1})}(z)$, where, $\mathcal{Q}_{(\mathcal{H}_n \mathcal{H}_{n+1})}(z) := \mathbb{P}(\bar{H}_{n+1} = \mathcal{H}_{n+1} \mid \bar{H}_{n+1} = \mathcal{H}_n, z_n)$.*

(b) *For all $z \in c^{|\mathcal{S} \times \boldsymbol{A}|} \times \Pi^\lambda$, the transition probabilities $\mathcal{Q}_{(\mathcal{H}_n \mathcal{H}_{n+1})}(z)$ form aperiodic, irreducible Markov chains over $\bar{H}$ and for all $s \in \mathcal{S}$ and $(s, \boldsymbol{a}) \in \mathcal{S} \times \boldsymbol{A}$, there exists $\mathcal{H}, \acute{\mathcal{H}} \in \bar{H}$, such that $s \in \mathcal{H}$ and $(s, \boldsymbol{a}) \in \acute{\mathcal{H}}$.*

(c) *The map $z \longmapsto \mathcal{Q}_{(\mathcal{H}_n \mathcal{H}_{n+1})}(z)$ is Lipschitz continuous.*

In effect, lemma III verifies that asymptotically every state of the game $\Gamma$ will be visited a minimum proportion of time, say $\tau > 0$. Also, the $\varepsilon$-trembled action transition probabilities in (32) ensures that every joint action will be selected with a non-zero probability; hence, every state-action pair is used a minimum proportion of time, say $\acute{t} > 0$.

**Lemma IV.** *Given any norm $\|.\|$ on $\mathbb{R}^{|\mathcal{S} \times \boldsymbol{A}|}$ and on $\mathbb{R}^{|\mathcal{S}|}$, there exists constants* A,B,C, *and* D *such that:*

$$\mathbb{E}\left[\left(V_{k,(s,\boldsymbol{a})}^n\right)^2 \Big| \mathcal{F}_n\right] < \mathrm{A} + \mathrm{B}\left\|\hat{Q}_{k,(s,\boldsymbol{a})}^n\right\|^2, \quad and \quad \mathbb{E}\left[\left(U_{k,s}^n\right)^2 \Big| \mathcal{F}_n\right] < \mathrm{C} + \mathrm{D}\|\boldsymbol{\pi}_s^n\|^2 \; \forall s, \boldsymbol{a}.$$

### C. Differential Inclusion Arguments

In this section, we proceed to characterize the limiting behavior of CNR$Q$-learning using differential inclusion arguments from [25]. Methodologically, the arguments in [25] are based on the well-established ODE approach [31] which treats the stochastic approximation (22) as a noisy discretization of an autonomous ODE with $F(x)$ as its mean-field. More specifically, under appropriate conditions on the step-sizes, mean-field and noise components of (22), it follows that the continuous-time linear interpolation of $x_n$ asymptotically tracks the stable fixed points of the dynamical system $\dot{x} = F(x)$. Hence, the limit sets of (22) will coincide with the set of stable fixed points of its associated ODE, and one can study instead the stability of the deterministic system $\dot{x} = F(x)$ to establish the convergence of the random sequence $\{x^n\}_{n \in \mathbb{N}}$. The results in [25] extend the ODE method to the case of asynchronous stochastic approximation

 

with set-valued mean-fields. Within this perspective, our next lemma (Lemma V) characterizes the limiting behavior of the fast stochastic recursion in (34). Before stating the lemma, we briefly hint on the main theoretical result in [25] which considerably facilitates our analysis in this paper. Our overview here is merely to convey the key idea in [25] in non-technical terms, and an avid reader is encouraged to consult [25] for a more technical exposition.

In case we were dealing with a synchronous updating pattern in CNR$Q$, standard arguments (e.g., [31]) would suggest that we may analyze the convergence behavior of the discrete-time iterates $\{\boldsymbol{\pi}^n\}_{n \in \mathbb{N}}$ by studying the limit sets of an associated ordinary differential inclusion (ODI) with correspondence $G(.,.)$ as its mean field:

$$\frac{d\boldsymbol{\pi}^t}{dt} \in G(\hat{Q}_k, \boldsymbol{\pi}^t),$$

where $\{\hat{Q}_k^n\}_{n \in \mathbb{N}}$ iterates on the slow timescale are frozen at $\hat{Q}_k$ by standard multi-timescale results [32]. However, unlike synchronous stochastic approximation where the steps sizes $\alpha(n)$ are deterministic, CNR$Q$ features random and time-varying step sizes of the form $\gamma(\phi^n(s)).\mathbb{I}_{\{s=s^n\}}$. The conventional approach to dealing with such asynchronicity does not immediately extend to set-valued mean fields; also, even if $G(.,.)$ were single-valued, the standard procedure would be to study the limit sets of a non-autonomous ODE of the form [31]:

$$\frac{d\boldsymbol{\pi}^t}{dt} = M(t).G(\hat{Q}_k, \boldsymbol{\pi}^t), \quad (41)$$

where, $M(.)$ is a matrix-valued measurable process such that $M(t)$ for each $t$ is a diagonal matrix with non-negative diagonal entries, reflecting the relative instantaneous rates with which the different components of $\boldsymbol{\pi}$ get updated. The existing theory does not explicitly define the scaling matrix $M(.)$ and it is further assumed that in the limit all the components of $\boldsymbol{\pi}$ are updated in an equally spaced manner and some 'specific' minimum proportion of the iterations. To work around the difficulties in studying (41), the approach in [25] shows that under the conditions stated in Lemma(s) I to IV, the diagonal elements of $M(t)$ lie almost surely in the closed set $[\tau, 1]$, for some $\tau > 0$. It then combines the set $[\tau, 1]$ with the mean field $G(.,.)$ to form a set-valued mean-field. More specifically, let $\Omega_{|\mathcal{S}|}^{\tau}$ be the $|\mathcal{S}| \times |\mathcal{S}|$ diagonal matrix of the form: $\Omega_{|\mathcal{S}|}^{\tau} := \{\mathrm{diag}(\xi_1, \dots, \xi_{|\mathcal{S}|}); \xi_s \in [\tau, 1], \forall s \in \mathcal{S}\}$. It is shown in [25] that the limit set of the asynchronous iterates $\{\boldsymbol{\pi}^n\}_{n \in \mathbb{N}}$ can be characterized by the asymptotic analysis of the following ODI:

$$\frac{d\boldsymbol{\pi}^t}{dt} \in \Omega_{|\mathcal{S}|}^{\tau}.G(\hat{Q}_k, \boldsymbol{\pi}^t).$$

This procedure pays off in two ways: first, the analysis can be done by studying an 'autonomous' rather than a 'non-autonomous' system; second, it extends the previous theory to also capture the behavior of asynchronous updates with 'set-valued' mean fields. Moreover, as argued in [25], it only suffices to verify that $\tau$ is positive; i.e., to ensure that all the components of the iterates get updated some minimum proportion of time. The key advantage lies in that the exact value of $\tau$ does not need to be known, as the analysis will be conducted for every $\tau > 0$. Now, recall from Lemma III that every $s \in \mathcal{S}$ in CNR$Q$ is, in fact, selected some minimum proportion of time, $\tau > 0$.

Armed with this understanding, we are now prepared to state Lemma V which corresponds to ([25], Assumption (B6)). Define the correspondence $\Pi^{ce}(.): c^{|\mathcal{S} \times A|} \mapsto \Pi^{\lambda}$ such that for all $\hat{Q}_k \in c^{|\mathcal{S} \times A|}$, one has $\boldsymbol{\pi} \in \Pi^{\lambda}$ is in $\Pi^{ce}(\hat{Q}_k)$ if and only if it satisfies Definition 1 for CE policies. $\Pi^{ce}(.)$ is an upper semi-continuous set-valued map (c.f, [15],

Lemma 14), such that for all $\hat{Q}_k \in c^{|\mathcal{S} \times A|}$, $\Pi^{ce}(\hat{Q}_k)$ is compact, convex, and non-empty (c.f., [15], Lemma 16).

**Lemma V.** *For all $\hat{Q}_k \in c^{|\mathcal{S} \times A|}$,*

*(a) the differential inclusion:*

$$\dot{\boldsymbol{\pi}}_s^t = \frac{d\boldsymbol{\pi}_s^t}{dt}$$
$$\in \Omega^\tau . G_s(\hat{Q}_k, \boldsymbol{\pi}_s^t), \quad \text{for all } s \in \mathcal{S}, \tag{42}$$

*is globally attracted by $\Pi^{ce}(\hat{Q}_k)$.*

*(b) $F(\hat{Q}_k, \Pi^{ce}(\hat{Q}_k))$ is a convex map.*

**Proof.** Following [16],[17], the correspondence $\Pi^{ce}(\hat{Q}_k)$ coincides with the set of no-regret policies; i.e.,

$$\{\boldsymbol{\pi} \in \Pi^\lambda \colon \mathcal{R}_{k,\langle a,\acute{a}\rangle,s}(\hat{Q}_k, \boldsymbol{\pi}) \leq 0, \forall k \in \mathcal{K}, \forall s \in \mathcal{S}, \forall a, \acute{a} \in A_k\}, \tag{43}$$

where, for general $\hat{Q}_k \in c^{|\mathcal{S} \times A|}$ and $\boldsymbol{\pi} \in \Pi^\lambda$, $\mathcal{R}_{k,\langle a,\acute{a}\rangle,s}$ is defined as:

$$\mathcal{R}_{k,\langle a,\acute{a}\rangle,s}(\hat{Q}_k, \boldsymbol{\pi}) \stackrel{\text{def}}{=} \sum_{\boldsymbol{a} \in A: a_k = a} \boldsymbol{\pi}_s(\boldsymbol{a}) . \left[\hat{Q}_{k,(s,\acute{a},\boldsymbol{a}_{-k})} - \hat{Q}_{k,(s,\boldsymbol{a},\boldsymbol{a}_{-k})}\right], \quad \forall k \in \mathcal{K}, \forall s \in \mathcal{S}, \forall a, \acute{a} \in A_k. \tag{44}$$

Equation (43) implies that the solutions to (42) steer the state-dependent regret matrix $R_{k,s}$ to approach the closed negative orthant $\mathbb{R}_{-}^{|A_k \times A_k|}$, denoted for short by $\Theta$. The analysis would be more direct if we consider the equivalent dynamics in the regret space; i.e., to show that for the solutions to (45) below:

$$\dot{R}_{k,s}^t = \frac{dR_{k,s}^t}{dt} \in \Omega^\tau . \left[\mathcal{R}_{k,s}\left(\hat{Q}_k, \Psi_s(\hat{Q}_k, \boldsymbol{\pi}^t)\right) - R_{k,s}^t\right], \tag{45}$$

we have that:

$$R_{k,\langle a,\acute{a}\rangle,s}^t \to \Theta \quad as \quad t \to \infty, \qquad \forall k \in \mathcal{K}, \forall s \in \mathcal{S}, \forall a, \acute{a} \in A_k.$$

Following ([25], Theorem 5.2), we now produce a Lyapunov function for (45) to show that $\Theta$ (resp. $\Pi^{ce}$) is a global attractor for (45) (resp. (25)). Define:

$$\mathcal{L}(R_k) = \frac{1}{2} \sum_{s \in \mathcal{S}, a, \acute{a} \in A_k} \left[\Upsilon\left(R_{k,\langle a,\acute{a}\rangle,s}\right)\right]^2.$$

Clearly, $\mathcal{L} \geq 0$, $\mathcal{L}(\Theta^{|\mathcal{S}|}) = 0$, and $\nabla \mathcal{L}(R_k) = \Upsilon(R_k)$. To show that $\mathcal{L}$ is a Lyapunov function for the ODI (28), one needs to verify for any fixed $\omega_s \in \Omega^\tau$ and any $\acute{\boldsymbol{\pi}} \in \Psi_s(\hat{Q}_k, \boldsymbol{\pi}^t)$:

$$< \nabla \mathcal{L}(R_k), \dot{R}_k^t > = \sum_{s \in \mathcal{S}} < \Upsilon(R_{k,s}), \omega_s . \left[\mathcal{R}_{k,s}(\hat{Q}_k, \acute{\boldsymbol{\pi}}) - R_{k,s}\right] > < 0, \quad \text{for } \forall R_{k,s} \in \mathbb{R}^{|A_k \times A_k|} \backslash \Theta \tag{46}$$

where $<.,.>$ denotes the Frobenius inner product. It can be shown (c.f., Lemma B.1 in the supplementary materials of the paper) that $< \Upsilon(R_{k,s}), \mathcal{R}_{k,s}(\hat{Q}_k, \acute{\boldsymbol{\pi}}) > = 0$ for all $s \in \mathcal{S}$; hence, (46) reduces to: $-\sum_{s \in \mathcal{S}} < \Upsilon(R_{k,s}), \omega_s . R_{k,s} >$ which is clearly less than 0. This concludes part (a).

As for part (b), the convexity of $F(\hat{Q}_k, \Pi^{ce}(\hat{Q}_k))$ is immediate given that the function $\mathrm{H}^{\lambda_k}$ in the definition of $F$ is an affine function of $\mathcal{L}_k^{\lambda_k}$, and for any fixed $\hat{Q}_k \in c^{|\mathcal{S} \times A|}$, $\mathcal{L}_k^{\lambda_k}$ reduces to a linear function of $\boldsymbol{\pi} \in \Pi^\lambda$. ∎

Now remember from Lemma III that every state-action pair is used a minimum proportion of time, $\acute{t} > 0$. Define $\Omega_{|A|}^{\acute{t}}$ to be the $|A| \times |A|$ diagonal matrix of the form: $\Omega_{|A|}^{\acute{t}} := \{\zeta_1, \ldots, \zeta_{|A|}; \zeta_{\boldsymbol{a}} \in [\acute{t}, 1], \forall \boldsymbol{a} \in A\}$. In light of ([25], Theorem 4.7), under assumption (A1) and with Lemmas I to V holding, the linear interpolation of the iterative process in (33) is an asymptotic pseudo-trajectory to the differential inclusion:

$$\frac{d\hat{Q}_{k,s}^t}{dt} \in \Omega_{|\boldsymbol{A}|}^{\dagger} . F_s\left(\hat{Q}_k^t, \Pi_{ce}\right), \quad \text{for all } s \in \mathcal{S} \quad (47)$$

where $F_s$ stands for the $|\boldsymbol{A}|$-vector of the $F_{(s,\boldsymbol{a})}$ terms; i.e., for any $\boldsymbol{\pi}^{ce} \in \Pi^{ce}$,

$$F_s\left(\hat{Q}_k^t, \boldsymbol{\pi}^{ce}\right) = \mathrm{H}_s^{\lambda_k}\left(\hat{Q}_k^t, \boldsymbol{\pi}^{ce}\right) - \hat{Q}_{k,s}^t, \quad \text{for all } s \in \mathcal{S},$$

and $\mathrm{H}_s^{\lambda_k}\left(\hat{Q}_k^t, \boldsymbol{\pi}^{ce}\right)$ is the $|\boldsymbol{A}|$-vector of $\mathrm{H}_{(s,\boldsymbol{a})}^{\lambda_k}\left(\hat{Q}_k^t, \boldsymbol{\pi}^{ce}\right)$ terms, defined in (37).

The next lemma establishes the convergence of the $\{\hat{Q}_k^n\}_{n \in \mathbb{N}}$ iterates on the moderate time-scale.

**Lemma VI.** $Q_{k,s}^{\lambda_k}$ (defined in (11)) is the unique global attractor of the differential inclusion (47).

*Proof.* Clearly, $\mathrm{H}^{\lambda_k}\left(Q_k^{\lambda_k}, \boldsymbol{\pi}\right) = Q_k^{\lambda_k}$. Also, for any fixed $\boldsymbol{\pi} \in \Pi^{\lambda}$, $\mathrm{H}_s^{\lambda_k}\left(\hat{Q}_k, \boldsymbol{\pi}\right)$ is a contraction mapping w.r.t. sup norm $\|.\|_{\infty}$ (e.g., see [22]); i.e.,

$$\left\|\mathrm{H}_s^{\lambda_k}(\hat{Q}_k, \boldsymbol{\pi}) - \mathrm{H}_s^{\lambda_k}\left(\hat{\hat{Q}}_k, \boldsymbol{\pi}\right)\right\|_{\infty} \le \rho. \left\|\hat{Q}_{k,s} - \hat{\hat{Q}}_{k,s}\right\|_{\infty}, \quad \forall \hat{Q}_k, \hat{\hat{Q}}_k \in c^{|\mathcal{S} \times \boldsymbol{A}|}.$$

which means that $Q_{k,s}^{\lambda_k}$ is its unique fixed point. Now, for any fixed $\omega \in \Omega_{|\boldsymbol{A}|}^{\dagger}$, we have:

$$\omega. \left(\mathrm{H}_s^{\lambda_k}(\hat{Q}_k, \boldsymbol{\pi}) - \hat{Q}_{k,s}\right) = \mathrm{H}_s^{\lambda_k, \omega}(\hat{Q}_k, \boldsymbol{\pi}) - \hat{Q}_{k,s},$$

where, $\mathrm{H}_s^{\lambda_k, \omega}(.) \stackrel{\text{def}}{=} (\mathbf{I} - \omega). \hat{Q}_{k,s} + \omega. \mathrm{H}_s^{\lambda_k}(\hat{Q}_k, \boldsymbol{\pi})$. Since $\omega$'s diagonal elements are bounded by 1, it holds that:

$$\left\|\mathrm{H}_s^{\lambda_k, \omega}(\hat{Q}_k, \boldsymbol{\pi}) - \mathrm{H}_s^{\lambda_k, \omega}\left(\hat{\hat{Q}}_k, \boldsymbol{\pi}\right)\right\|_{\infty} \le \bar{\rho}. \left\|\hat{Q}_{k,s} - \hat{\hat{Q}}_{k,s}\right\|_{\infty}, \quad \forall \hat{Q}_k, \hat{\hat{Q}}_k \in c^{|\mathcal{S} \times \boldsymbol{A}|},$$

where $\bar{\rho} \stackrel{\text{def}}{=} 1 - \zeta^*(1 - \rho) \in (0,1)$, and $\zeta^* = \max_i \zeta_i$. Thus, $\mathrm{H}_s^{\lambda_k, \omega}(\hat{Q}_k, \boldsymbol{\pi})$ is also a contraction mapping, and $Q_{k,s}^{\lambda_k}$ is its unique fixed point; i.e., $\mathrm{H}^{\lambda_k, \omega}\left(Q_k^{\lambda_k}, \boldsymbol{\pi}\right) = Q_k^{\lambda_k}$. From this, it follows that $\{\hat{Q}_k^n\}_{n \in \mathbb{N}}$ converge to true action values $Q_k^{\lambda_k}$ for any policy $\boldsymbol{\pi} \in \Pi^{\lambda}$, and in particular for the CE policies in $\Pi^{ce}$. ■

**Theorem II.** The coupled process $\left(\hat{Q}_k^n, \boldsymbol{\pi}^n\right)$ from (33) and (34) converges to the limit $\left(Q_k^{\lambda_k}, \Pi^{ce}\left(Q_k^{\lambda_k}\right)\right)$, where $\boldsymbol{\pi}^{ce} \in \Pi^{ce}\left(Q_k^{\lambda_k}\right)$ is a stationary CE policy for the stochastic game $\Gamma^{\lambda}$ with $\lambda_k$-parameterized individual Lagrangian utilities and $Q_k^{\lambda_k}$ is the associated Lagrangian state-action value function.

*Proof.* Immediate by Lemma VI and ([25], Corollary 4.8). ■

Theorem II establishes the convergence of $\{\boldsymbol{\pi}^n\}_{n \in \mathbb{N}}$ to a small neighborhood of the set of CE in game $\Gamma^{\lambda}$. Hence, by Theorem I, $\mathcal{L}_k^{\lambda_k}\left(\hat{Q}_k^n, \boldsymbol{\pi}^n\right)$ converges to the Lagrange dual function $\mathcal{G}_k(\lambda_k)$ that is equal to the primal maximum $\check{\mathcal{L}}_{k,\check{s}_k}^{\lambda_k}(\check{\pi}_k^*)$ in view of the agents' individual control problems in section II.A. With this equivalence in mind, the convergence of $\{\lambda_k^n\}_{n \in \mathbb{N}}$ to the set of dual minima $\lambda_k^* \in \arg\min_{\lambda_k} \mathcal{G}_{k,s}(\lambda_k)$ can be established similarly to [33]. More specifically, the mapping $\lambda_k \to \mathcal{G}_{k,s}(\lambda_k)$ is piecewise linear and convex for every state $s \in \mathcal{S}$ and any agent $k$. Let $\nabla_{\lambda_k}$ denote the gradient in the $\lambda_k$ variable. We have $\nabla_{\lambda_k}\mathcal{G}_{k,s}(\lambda_k^t) \in \partial \mathcal{G}_{k,s}(\lambda_k^t)$, where $\partial \mathcal{G}$ is the sub-differential of $\mathcal{G}$. It then holds that [33] the stochastic (sub-)gradient iterations on $\{\lambda_k^n\}_{n \in \mathbb{N}}$ track the differential inclusion:

$$-\dot{\lambda}_k^t \in \partial \mathcal{G}_{k,s}(\lambda_k^t) \quad \text{for all } s \in \mathcal{S},$$

and therefore converges to the set of minima of $\mathcal{G}_{k,s}$ [33][33]. Combining this with Theorem II yields that $\left(\hat{Q}_k^n, \boldsymbol{\pi}^n, \lambda_k^n\right)$ converge to $\left(Q_k^{\lambda_k^*}, \Pi^{ce}\left(Q_k^{\lambda_k^*}\right), \lambda_k^*\right)$, which is an alternative way of saying $\mathcal{L}_k^{\lambda_k}(\hat{Q}_k^n, \boldsymbol{\pi}^n)$ converges to $\check{\mathcal{L}}_k^{\lambda_k^*}(\check{\pi}_k^*)$ associated with the saddle point of the individual agent's long-term Lagrangian simultaneously for all agents $k \in \mathcal{K}$.

# V. NUMERICAL RESULTS

In this section, we follow up on the example scenarios from Section II.C, and give numerical results on CNR$Q$ learning algorithm. We conduct experiments for a two-tier network with four femto-cells and a single macro-cell. We investigate CNR$Q$'s convergence and also compare its social welfare with that computed from both the centralized and semi-distributed variants of the CE-$Q$ learning algorithm both implemented with a utilitarian equilibrium selection mechanism [20]. In order to apply CE-$Q$ to our constrained game example, we have adopted the Lagrangian approach similarly to CNR$Q$ and have augmented CE-$Q$ with Lagrange multiplier iterations that run on a slower timescale w.r.t. $Q$-value iterations. Since centralized CE-$Q$ is convergent to stationary CE, this would result in proper handling of the constraints in the game; however, semi-distributed CE-$Q$ is susceptible to mis-coordination, and convergence to CE is not guaranteed in general. This is also corroborated by our experiments in the downlink scenario in that semi-distributed CE-$Q$ violates the constraint on MBS's buffer length.

### TABLE IV
### SIMULATION PARAMETERS- UPLINK SCENARIO

| Symbol | Quantity | Value |
|---|---|---|
| $No$ | noise power [mW] | $10^{-7}$ |
| $a_0^u$ | MUE transmit power [mW] | 5 |
| $a_k^u, k = 1 \dots 4$ | FUE transmit power [mW] | $\{'low' \cong 0,' high' \cong 1\}$ |
| $g_{0k}^u, k = 1 \dots 4$ | MUE-FBS channel gains | $(0.038, 0.082, 0.071, 0.086)$ |
| $h_{k\acute{k}}^u, k, \acute{k} = 1 \dots 4$ | FUE-FBS channel gains | $\begin{pmatrix} 0.44 & 0.10 & 0.02 & 0.10 \\ 0.07 & 0.23 & 0.03 & 0.06 \\ 0.10 & 0.10 & 0.25 & 0.10 \\ 0.10 & 0.05 & 0.09 & 0.24 \end{pmatrix}$ |
| $W$ | bandwidth [MHz] | 1 |
| $\bar{a}_k^u, k = 1 \dots 4$ | mean FUE power constraint [mW] | 0.75 |

### TABLE V
### SIMULATION PARAMETERS- DOWNLINK SCENARIO

| Symbol | Quantity | Value |
|---|---|---|
| $a_0^d$ | MBS transmit power [mW] | 500 |
| $a_k^d, k = 1 \dots 4$ | FBS transmit power [mW] | $\{0, 10, 100\}$ |
| $g_{0k}^d, k = 1 \dots 4$ | MBS-FUE channel gains | $(0.003, 0.005, 0.008, 0.002)$ |
| $g_{k0}^d, k = 1 \dots 4$ | FBS-MUE channel gains | $(0.055, 0.051, 0.035, 0.012)$ |
| $h_{k\acute{k}}^d, k, \acute{k} = 1 \dots 4$ | FBS-FUE channel gains | $\begin{pmatrix} 0.68 & 0.09 & 0.03 & 0.04 \\ 0.07 & 0.82 & 0.04 & 0.04 \\ 0.01 & 0.04 & 0.16 & 0.03 \\ 0.03 & 0.08 & 0.01 & 0.29 \end{pmatrix}$ |
| $\tau$ | timeslot duration [msec] | 1 |
| $\lambda$ | MBS-to-MUE Poisson packet arrival rate [pkt/msec] | 5.5 |
| $L$ | packet size [bytes] | 256 |
| $\bar{b}_0$ | mean MBS buffer length constraint [pkt] | 10 |

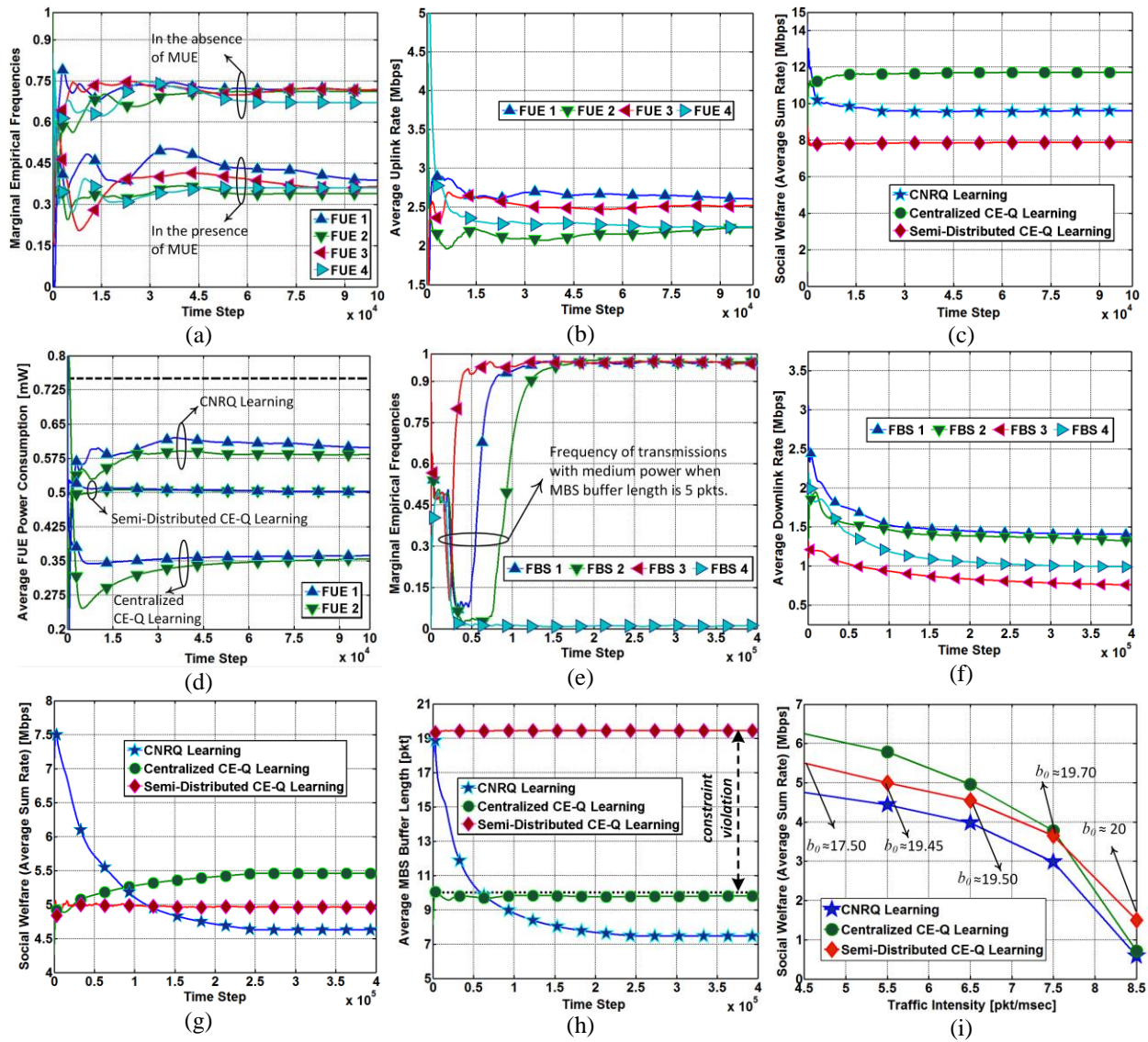

Fig. 2. (a) FUEs' marginal frequency of transmissions (uplink). (b) Individual FUE's average rate utilities. (c) Average uplink social welfare. (d) Convergence of individual FUE's average power consumption. (e) FBSs' marginal frequency of transmissions (downlink). (f) Individual FBS's average rate utilities. (g) Average downlink social welfare. (h) Convergence of MBS's average buffer length. (i) Downlink social welfare vs. MBS traffic intensity (note the violation of constraint in semi-distributed CE-$Q$ learning).

First, consider the uplink HetNet setup from II.C.1. The simulation parameters are listed in Table IV. Fig. 2(a) and (b) exhibit the convergence behavior of CNR$Q$ in this scenario. In Fig. 2(a), we plot the marginal empirical frequency of high power transmissions (action $a_k^u = 1$) by FUEs for both cases of MUE's occupancy state. In Fig. 2(b), the progression of the average individual rate utility achieved by all FUEs is depicted. In Fig. 2(c), we compare CNR$Q$'s social welfare (measured in terms of the sum of FUEs rate utilities) with that obtained from both semi-distributed and centralized versions of the CE-$Q$ algorithm. As can be seen, CNR$Q$ outperforms semi-distributed CE-$Q$, but its social welfare is upper bounded by centralized CE-$Q$. We show in Fig. 2(d) the average power consumption by FUEs. To keep the figure from being cluttered, the results are shown only for FUEs 1 and 2. The imposed average power constraint (0.75 mW in Table IV) is respected asymptotically by all three algorithms.

To experiment with the downlink setup, we use the simulation parameters listed in Table V. Fig. 2(e) shows the convergence of marginal empirical frequency of play for action $a_\kappa^d = 10\,[mW]$ by each FBS when the system state (i.e., MBS buffer length) is $b_0 = 5$. Fig. 2(f) depicts the convergence of the average rate achieved by each individual FBS. In Fig. 2(g) and (h), we compare CNR$Q$'s social welfare and constraint satisfaction with the other two schemes. As evidenced, semi-distributed CE-$Q$, despite achieving a slightly higher average sum rate, has violated the constraint on average MBS buffer length by a relatively large margin. We also study the impact of the MBS's Poisson traffic arrival rate on the downlink social welfare and on the constraint on MBS buffer length. To this end, the traffic intensity is varied from 4.5 pkt/msec to 8.5 pkt/msec. MBS buffer length constraint is consistently respected by both CNR$Q$ and centralized CE-$Q$. However, as shown in Fig. 2(i), despite its high social welfare, semi-distributed CE-$Q$ has consistently violated the constraint on buffer length.

## VI. CONCLUSIONS

We presented a constrained no-regret $Q$-learning (CNR$Q$) algorithm for the online computation of stationary CE in constrained general-sum stochastic games. CNR$Q$ builds on previous ideas which involve two control loops consisting of $Q$-learning (outer-loop) for estimating action value functions and no-regret-learning (inner-loop) for estimating a CE policy. We employed the technique of timescale separation from stochastic approximation to allow for a single-loop concurrent execution of $Q$-learning (on the slower timescale) and no-regret-learning (on the faster timescale), which eliminates the backstage virtual plays as required by prior art in inner-loop iterations. Moreover, by regarding distributed CE estimation as simultaneous primal maximization across all agents, we extended the algorithm for constrained setups as well. Thanks to our stochastic approximation-based expression of the learning process, the constrained extension comes as easily as introducing a slower third timescale to the operation of the algorithm for conducting dual descent in Lagrange multiplier space. Overall, CNR$Q$ has been cast as a three-timescale asynchronous stochastic approximation with set-valued update increments. Unlike prior art which lacks a rigorous convergence analysis, we analyzed the asymptotic behavior of CNR$Q$ using differential inclusion arguments. Our analysis draws on recent extensions of the theory of stochastic approximation to the case of asynchronous recursive inclusions with set-valued mean fields. We also applied CNR$Q$-learning to an exemplary case of emerging wireless HetNet deployments.